\documentclass[10pt,journal,compsoc]{IEEEtran}

\usepackage{graphicx}
\usepackage{amssymb,amsmath,amsbsy,epsfig,float,lscape,makeidx}
\usepackage{algorithm}
\usepackage{algorithmicx}
\usepackage[shortlabels]{enumitem}
\usepackage{subfig}
\usepackage{tikz}
\usepackage{comment}
\usepackage{url}
\usepackage{soul}

\ifCLASSOPTIONcompsoc
  \usepackage[nocompress]{cite}
\else
  \usepackage{cite}
\fi

\def\argmin{\mathop{\rm argmin}}

\newcommand{\real}{\ensuremath{\mathbb{R}}}

\newcommand{\ltwo}{\ensuremath{\mathbb{L}^2}}

\newcommand{\shapes}{{\cal S}}
\newcommand{\W}{{\cal W}}

\newcommand{\R}{\mathbb{R}}

\usepackage{color}
\usepackage{amsthm}

\newtheorem{proposition}{Proposition}
\newtheorem{theorem}{Theorem}
\newtheorem{definition}{Definition}
\newtheorem{remark}{Remark}

\begin{document}

\title{Statistical Shape Analysis of Shape Graphs with Applications to Retinal Blood-Vessel Networks}

\author{Aditi~Basu~Bal,
        Xiaoyang~Guo,
        Tom~Needham, 
        and~Anuj~Srivastava
\IEEEcompsocitemizethanks{\IEEEcompsocthanksitem A.  B. Bal and A. Srivastava are at the Department
of Statistics and T. Needham is at the Department of Mathematics, Florida State University, Tallahassee, FL 32306. \protect\\
E-mail: ab18z@fsu.edu
\IEEEcompsocthanksitem X. Guo is working at Meta.}
}

\IEEEtitleabstractindextext{%
\begin{abstract}
This paper provides theoretical and computational developments in statistical shape analysis of 
{\it shape graphs}, and demonstrates them using analysis of complex data from retinal blood-vessel (RBV) networks.
The shape graphs are represented by a set of nodes and  
edges (planar articulated curves) connecting some of these nodes. The goals are to utilize shapes of 
edges and connectivities and locations of nodes to: (1) characterize full shapes, (2) quantify
shape differences, and (3) model statistical variability. We develop a mathematical representation, elastic Riemannian shape metrics, and associated tools for such statistical analysis. Specifically, we derive tools for shape graph registration,  geodesics, summaries, and shape modeling. Geodesics are convenient for visualizing optimal deformations, and PCA helps in dimension reduction and statistical modeling. 
One key challenge here is comparisons of shape graphs with vastly different complexities (in number of nodes and edges). This paper introduces a novel multi-scale representation of shape graphs to handle this challenge. Using the notions of (1) ``effective resistance" to cluster nodes and (2) elastic shape averaging of 
edge curves, one can reduce shape graph complexity while maintaining overall structures. This way, we can compare shape graphs 
by bringing them to similar complexity.
We demonstrate these ideas on Retinal Blood Vessel (RBV) networks taken from the STARE and DRIVE databases.
\end{abstract}

}

\maketitle

\IEEEdisplaynontitleabstractindextext

%
\IEEEpeerreviewmaketitle

\IEEEraisesectionheading{\section{Introduction}\label{sec:introduction}}

\IEEEPARstart{I}{maging} is a rich data source for studying {\it structures} found in diverse biological, ecological, bioinformatical, and anatomical systems. Due to rapid advances in imaging resolutions and processing techniques, this data increasingly involves objects with complex morphologies, resulting in growing challenges for the shape analysis community. While early efforts focused on shape analysis of point sets, curves, or surfaces, the focus has shifted to objects with branching and network structures, such as trees and graphs. Trees are mathematical objects admitting a hierarchical organization as a set of nodes connected by edges (or branches). A primary branch connects to secondary branches, those, in turn, connect to the tertiary branches, and so on. Shape graphs are relatively less structured than trees. A {\it  shape graph is a finite collection of nodes with some subset of nodes connected in a pairwise manner by curves}. There is usually no hierarchy of nodes or edges in shape graphs. The shape of such a graph incorporates the placements and connectivities of nodes and the shapes of edges. Note that the term {\it graph} is generally used for a set of nodes connected by edges with Euclidean attributes; we will distinguish our data objects, where the edges are shapes of articulated curves, by calling them {\it shape graphs}. 

\begin{figure}
    \centering
    \includegraphics[scale=0.2]{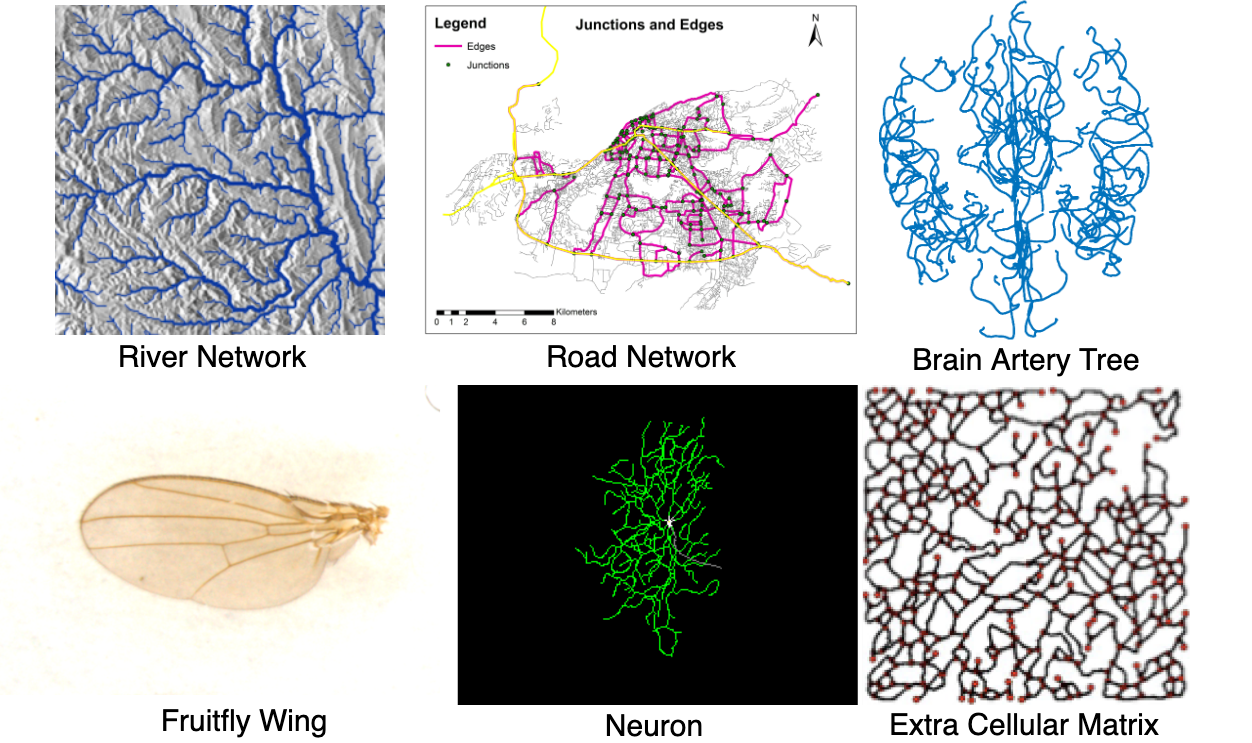}
    \caption{Shape networks in nature and biological structures: River Network (WWF HydroSHEDS Free Flowing Rivers Network v1), Road Network \cite{Das2019}, Brain Artery Tree, Fruitfly Wing, Neuron, Extracellular Matrix.}
    \label{fig:shape_graphs}
\end{figure}
There are two broad types of morphological data analyses -- feature-based and object-based. {\bf Feature-based} approaches extract some summary morphological features from the data and then analyze them in the feature space, often using Euclidean statistics. Although popular in the scientific community, this approach is limited because features contain only partial information about objects. These feature representations are typically non-invertible, i.e., one cannot go back from a feature vector to an object. {\bf Object-based} approaches are more comprehensive but cumbersome to implement. They rely on analysis in the original space of objects and provide statistical inferences as full objects. For instance, one can visualize shape averages or standard deviations around these averages as objects, which cannot be done using feature-based approaches. In this paper, we seek and develop an object-based strategy for analyzing shape graphs. 

Statistical shape analysis seeks to quantify, compare, model, and test shape populations from the observed data. The main challenge in analyzing shape graphs comes from their complexity. They are typically more complex in structure and variability than previously studied objects in the literature. There are very few, if any, object-based approaches in the literature for analyzing shape graphs. 
This is mainly due to the fact that object-based shape approaches require the registration of parts across objects when comparing two objects. This registration step is a big challenge in any shape comparison, more so for shape graphs. This paper develops metrics and procedures that extend past work on shape analysis of {\it elastic curves}, {\it elastic trees}, and {\it elastic surfaces} to {\it elastic shape graphs}.

\begin{figure}[h]
\begin{center}
\includegraphics[height=2.4in]{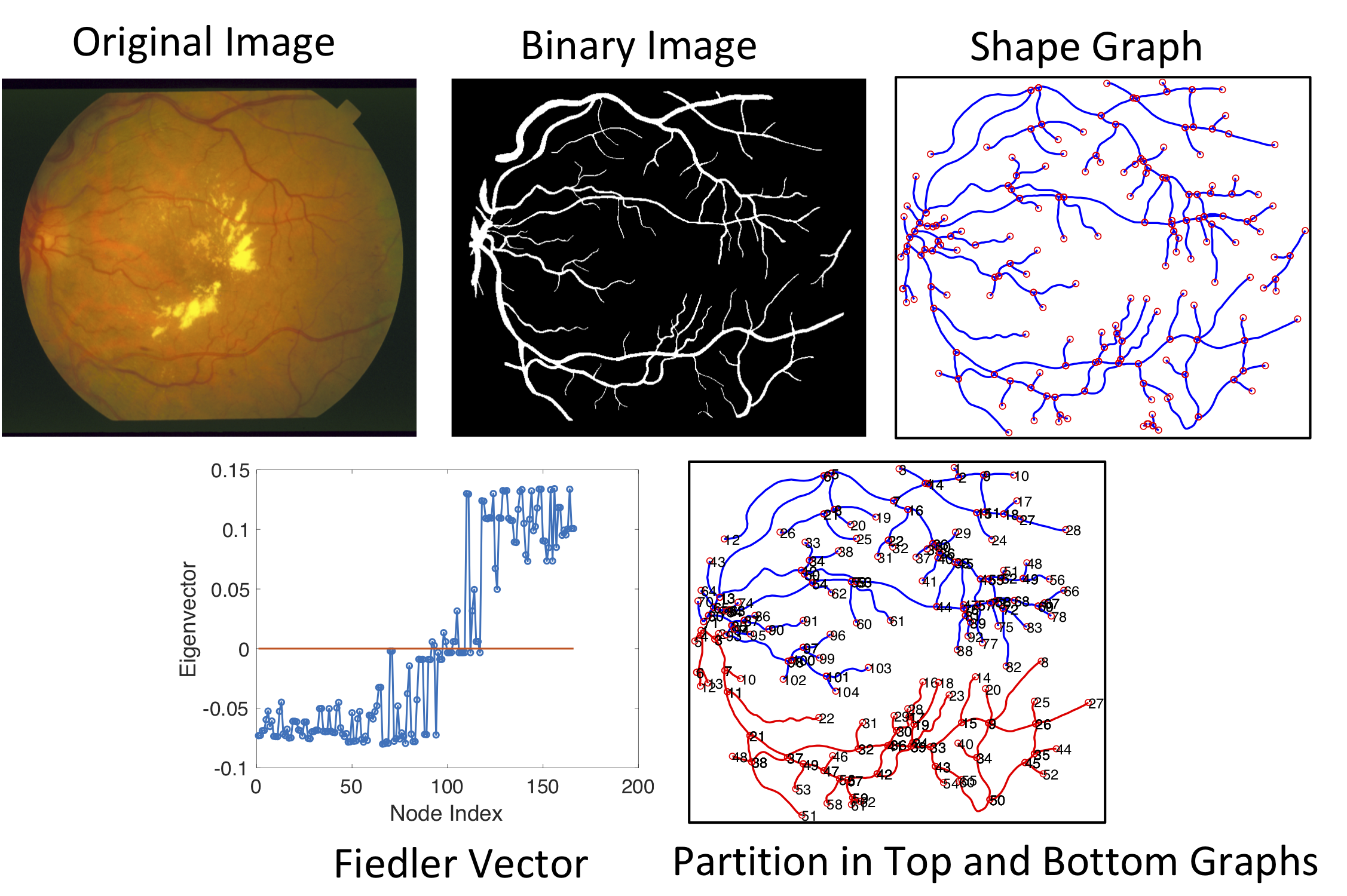}
\end{center}
\caption{Data Pre-Processing Pipeline. Images of 
RBVs (top-left) are first segmented into binary images (top-center) and then extracted as shape graph representation (top-right). These shape graphs are then partitioned into top and bottom components (bottom row).}
\label{fig:preprocess}
\end{figure}

We shall use both simulated and real shape graph data to demonstrate our approach. The real data correspond to the shapes of blood vessel networks
extracted from retinal images and taken from two publicly available databases STARE~\cite{hoover2000locating,STARE-project} and DRIVE~\cite{DRIVE-challenge}. We will call these shape graphs {\it Retinal Blood Vessel (RBV) networks}. 
An example of a retinal image is shown in the top row of Fig. \ref{fig:preprocess}. The retina is a light-sensitive layer of eye tissue that plays an essential role in visual perception. 
The retinal blood vessels (RBVs) supply oxygen and nutrition to guarantee the vitality of the retina and 
exhibit a rich and complex network structure, with branches and connections existing at arbitrary positions.
Understanding the structures of blood vessels is essential in 
understanding retinal functionality and diagnosing 
any degenerative diseases.

The problem of extracting RBVs from retinal images is challenging but not the focus of this paper. 
Currently, there exist several procedures, primarily based on deep learning, 
to handle the extraction problem.
In fact, for a subset of STARE data~\cite{hoover2000locating,STARE-project} and DRIVE data \cite{DRIVE-challenge} used in this paper, the authors also provide
manual segmentations and can be used directly. 
An example of binary segmentation is shown in the top row of Fig. \ref{fig:preprocess}.
We then apply the Vessel tech algorithm \cite{vessel-tech} on the binary image to extract the shape graph structure, {\it i.e.}, to determine the coordinates of nodes and points along the edges. 
This process may result in some small disconnected pieces, and 
we remove them using the {\it Depth-First Search} algorithm.
In summary, this process produces shape graphs that contain: (1) coordinates of points along the blood
vessels forming planar curves and (2) nodes where these vessels intersect, thereby forming a {\it shape graph}.
The top-right panel shows an example. 

To reduce the complexity of RBV shape graphs, we divide them into two parts: top and bottom. 
This division is performed using a simple graph-cut based on spectral clustering~\cite{ng2002spectral}.
Given an RBV network, we first compute its {\it binary} adjacency matrix and the associated Laplacian matrix, and 
use the {\it Fiedler vector} (the eigenvector of the Laplacian corresponding to the first nonzero eigenvalue) 
to partition the nodes into two groups -- top and bottom -- and study them as separate shape graphs. This process is illustrated in the bottom row of Fig. \ref{fig:preprocess}.
Some examples of the resulting RBV top and bottom shape graphs are shown later in Fig.~\ref{fig:clustering} and other results.
\\

\noindent {\bf Goals and Challenges}: Given such shape graph data, 
our goal is to analyze the full physical shapes of these objects, including the nodes, edges, and their connections.
Specifically, we seek to quantify shape differences across these objects, compute statistical shape summaries, 
and derive PCA-based Euclidean representations. We can then use these tools to develop 
statistical models and perform hypothesis testing for disease evaluations. 

The main challenges are the following. Any two shape graphs differ from each other in (1) the numbers of nodes and edges,
(2) the edge connectivities and node locations, and (3) the shapes of the edges formed by blood vessels. Together these factors imply that objects exhibit geometric and topological variability, making it challenging to compare shapes across entities. While this comparison is already tricky for objects with the same topology but different geometries, it becomes even more difficult for objects with different geometries and topologies. In the context of graphs, the matching of nodes is a combinatorial, NP-hard problem and one uses relaxation techniques to approximate
solutions. Our problem is more complex than the classical graph-matching, as we must also register points on edges across shape graphs. 
\\

\noindent {\bf Past Literature and Its Limitations}:  There are several research fields that are relevant to the current context. 
\\

\noindent -- {\bf Shapes of Curves}:
First, one needs tools for analyzing shapes of edges as curves -- we will utilize the elastic shape analysis approach here \cite{srivastava2016functional} but other ideas are also applicable (see e.g. \cite{younes-distance2}). 
\\

\noindent  -- {\bf Evolutionary or Phylogenetic Trees}: We note that there are many papers on analyzing tree shapes but considering only the branching structures, especially in phylogeny and evolutionary science~\cite{matsen-bio:2006,nye-ACM:2014,Chindelevitch-plos:2021}. These frameworks do not include shapes of branches in their analysis and thus are not entirely relevant to the current discussion. 
\\

\noindent -- {\bf Shape Trees}: There are several papers have extended such techniques to analyzing shapes of tree-like structures, using elastic analysis~\cite{wang2020statistical,duncan2018statistical,wang-laga-arXiv:2021}, LDDMM~\cite{Antonsanti-arXiv:2020}, or non-elastic shape analysis~\cite{feragon-etal-PAMI:2012}. As mentioned earlier, the hierarchy of nodes and a lack of loops in tree-like structures greatly simplifies the registration problem. It allows the use of linear assignment for registration of branches, instead of the quadratic assignment that is needed for graphs. 
\\

\noindent -- {\bf Networks or Graphs}: A growing body of research focuses solely on analyzing node connectivity in graphs through the lens of Riemannian geometry. Towards this goal, the use of quotient space structure for graph representations was introduced in \cite{jain2009structure} and extended in~\cite{calissano2020populations,guo-srivastava-diffcvml,guo2019quotient}. While these works compare graph structures by comparing node connections, they do not consider the shapes of edges in the graphs. Similarly, the Gromov-Wasserstein framework \cite{memoli2011gromov} has recently become a popular tool for studying graph data through probabilistic node correspondences \cite{xu2019scalable, vayer2020fused, chowdhury2021generalized, chowdhury2021quantized}. It enjoys a Riemannian structure \cite{sturm2012space}, allowing for the computation of geodesics between networks \cite{chowdhury2020gromov}. However, the literature on Gromov-Wasserstein distances has not been extended to the shape graph setting.  
\\

\noindent  -- {\bf Topological Data Analysis:} Tools from Topological Data Analysis (TDA) have been applied frequently to shape data \cite{chazal2009gromov, turner2014persistent, thomas2021topological}, including tree shapes \cite{bendich2016persistent}. Generally, these methods produce signatures that are amenable to statistical analysis. Still, the signatures cannot be inverted to enable direct inferences about the geometric features of the shapes. That is, TDA falls under the \emph{feature-based}, rather than \emph{object-based} paradigm.
\\

\noindent  -- {\bf Shape Graphs}: Finally, we mention past research that studies shape graphs in full generality.
Guo et al.~\cite{guo2020-BANs} was the first paper to formulate an analysis of shape graphs -- including node graphs and edge shapes. While this approach is comprehensive, the proposed solutions are relatively limited and less effective. It did not satisfactorily address the primary issue of matching nodes across large complex structures. Further, such essential statistical tools as shape means, PCA, and dimension reduction were treated in a limited way. More recently, \cite{Sukurdeep-arXiv-2021} provides a different mathematical framework for shape graph matching based on combining Riemannian structure with techniques from the theory of varifolds. However,  it does not offer any demonstrations of shape graphs beyond simple tree topologies. 
\\

\noindent {\bf Paper Contributions}:
The salient contributions of this paper are: 
\begin{enumerate}
\item {\bf Weighted Edge Representations}: From a theoretical perspective, this paper appends prior mathematical representations of \cite{guo2020-BANs} with additional {\it length} or weight variables for edge curves. This weighted representation allows for better interpretations and improvements in computing geodesic paths between shape graphs. These improvements are essential for producing interpretable statistical summaries of shape data.

\item {\bf Multi-Scale Shape Representations}: This paper introduces the notion of {\it multiscale} representations for shape graphs. 
It utilizes ideas from electrical circuit theory \cite{ellens2011effective}, along with elastic shape analysis of 
Euclidean curves \cite{srivastava2016functional}, in order to reach interpretable, scalable representations of complex shape graphs. This scalability allows us to study the shapes graphs at different levels of complexity. Specifically, it facilitates more natural comparisons of objects by bringing them to a similar scale (complexity).

\item {\bf Efficient Statistical Analysis}: It develops efficient, albeit approximate, alternatives to full-scale statistical analysis using low-resolution representations of shape graphs. It demonstrates the power of elastic analysis of shape graphs using handcrafted and RBV datasets.

\item {\bf Applications: RBV Network Shapes}: To the best of our knowledge, it is the first paper to provide a statistical shape analysis of complex graphs associated with RBV networks. Specifically, it provides tools for 
computing shape geodesics, shape summaries, and tangent PCA-based dimension reduction for RBV networks and their multiscale representations.
\end{enumerate}

\section{Modeling Shape Graphs}

This section outlines our precise mathematical representation of shape graphs, discussed intuitively above. We define a \emph{shape graph} in $\R^2$ to be the union of a finite collection of simple compact plane curves $\{C_i\}_{i=1}^m$; i.e., each $C_i$ is the image of a smooth immersion $[0,1] \to \R^2$. The \emph{endpoints} of such an immersion are the images of $0$ and $1$. We assume that the immersions are injective except, potentially, at their endpoints. We require that, for all $i \neq j$, $C_i \cap C_j$ is contained in the set of endpoints of the two curves (this includes the case that $C_i \cap C_j$ is empty). The set of intersection points is indexed with labels $V = \{v_i\}_{i=1}^n$ and we refer to these labels as \emph{nodes}. In analogy with classical graph theory terminology, we refer to each $C_i$ as an \emph{edge} of the shape graph and denote the set of edges as $E = \{C_i\}_{i=1}^m$. 

Due to their complex and widely varied topologies, shape graphs form a class of objects that are rather unwieldy from a shape analysis perspective. To put our analysis on firm footing, we introduce a model for shape graphs that borrows ideas from classical graph theory. This model extends the framework used in \cite{guo2020-BANs}, improving the model to give more intuitive interpretations of standard tools in shape analysis, such as geodesics and PCA. Our new formalism is best explained by starting with a more general class of attributed graphs, as originally studied in~\cite{jain2009structure}. 

\subsection{Metric Space of Attributed Graphs}\label{sec:metric_space_attributed_graphs}

Let $V$ be a fixed finite set of \emph{nodes} -- in this subsection, the nodes form an abstract indexing set and don't necessarily correspond to any shape graph -- and let $X = (X,d)$ be a fixed metric space. An \emph{$X$-attributed graph over $V$} is a function $A:V \times V \to X$, which, in analogy with the usual graph theory terminology, we refer to as an \emph{adjacency function}. One defines the distance between two such adjacency functions $A_0$ and $A_1$ as 
\begin{equation}\label{eqn:product_metric}
\left(\sum_{(v_i,v_j) \in V \times V} d(A_0(v_i,v_j),A_1(v_i,v_j))^2\right)^{\frac{1}{2}}.
\end{equation}
Observe that this metric on the space of adjacency functions is a familiar object -- it is simply the standard $\ell_2$-metric on the product space $X^{V \times V}$ (see, e.g.,~\cite[Section 3.6]{burago2001course}).

Recall that a metric space $(X,d)$ is called a \emph{geodesic space} if, for every pair of points $x_0,x_1 \in X$, there is a continuous path $x_u:[0,1] \to X$ such that for all $0 \leq s \leq t \leq 1$, $d(x_s,x_t) = (t-s)d(x_0,x_1)$; such a path is called a \emph{geodesic}. If $X$ is a geodesic space, then the space of adjacency functions valued in $X$ is also geodesic. Indeed, a geodesic between $A_0$ and $A_1$ is given by the path $A_u$ defined by setting $A_u(v,w)$ to be the geodesic in $X$ joining $A_0(v,w)$ to $A_1(v,w)$---this follows from the observation that our metric on the space of adjacency functions is just a product metric~\cite[Lemma 3.6.4]{burago2001course}. 

The more novel part of the theory developed in~\cite{jain2009structure} arises when comparing metric space-attributed graphs defined over \emph{different} node sets. Suppose that $A_i:V_i \times V_i \to X$ define $X$-attributed graphs for $i=0,1$. Assume for the sake of simplicity that $|V_0|=|V_1|=n$ and  write $V_0 = \{v^0_i\}_{i=1}^n$ and $V_1 = \{v^1_i\}_{i=1}^n$. One computes the distance between $A_0$ and $A_1$ by solving the \emph{graph matching problem}
\begin{equation}\label{eqn:graph_matching_metric_edges}
\min_{\sigma \in S_n} \left(\sum_{j=1}^n \sum_{i = 1}^n d\big(A_0(v^0_i,v^0_j),A_1(v^1_{\sigma(i)},v^1_{\sigma(j)})\big)^2\right)^{\frac{1}{2}},
\end{equation}
where $S_n$ is the group of permutations of $\{1,\ldots,n\}$. This optimization problem searches for matching between the node sets, which minimizes an overall distortion in attribute values. This idea extends to attributed graphs whose node sets have different cardinality. We delay such technical details until Section \ref{sec:computational_details}, where we will discuss them in the specific context of shape graphs.

Frequently, the attributed graphs are also endowed with node features. A \emph{node feature function} on a vertex set $V$ is a map $F:V \to Y$, where $(Y,D)$ is another metric space (more generally, $D$ can be any divergence function on $Y \times Y$). In this case, it makes sense to incorporate the node features into the comparison process. Suppose that we have $X$-valued adjacency functions $A_i:V_i \times V_i \to X$ and $Y$-valued node feature functions $F_i:V_i \to Y$ over different node sets $V_0$ and $V_1$ of the same cardinality. We then compute the distance between pairs $G^0 \equiv (A_0,F_0)$ and $G^1 \equiv (A_1,F_1)$ by solving the \emph{node-attributed graph matching problem}
\begin{align}\label{eqn:graph_matching_metric_edges_and_nodes}
& d_{\mathrm{graph}}(G^0, G^1) = \nonumber \\
& \min_{\sigma \in S_n} \left(\lambda\sum_{j=1}^n \sum_{i = 1}^n d\big(A_0(v^0_i,v^0_j),A_1(v^1_{\sigma(i)},v^1_{\sigma(j)})\big)^2 \right. \nonumber \\
&\qquad \qquad \qquad \qquad \left.+ (1-\lambda) \sum_{i=1}^n D\big(F_0(v^0_i),F_1(v^1_{\sigma(i)})\big)^2 \right)^{\frac{1}{2}},
\end{align}
where $\lambda \in [0,1]$ is a user-specified balance parameter. The minimum value is defined to be the distance between shape graphs and is used for comparing and summarizing shape graphs. We will use ${\cal G}$ to denote the set of all shape graphs each represented by a pair of functions $(A,F)$.

\subsection{Previous Shape Graph Model}

We now describe our previous mathematical model for shape graphs~\cite{guo2020-BANs}, which is based on the framework described in the previous subsection. Consider a shape graph $G=(V,E)$ with edges (curves) $E = \{C_i\}_{i=1}^m$ and nodes (labels for intersections of curves) $V = \{v_i\}_{i=1}^n$. 

It is natural to define a node feature function for $G$ as $F:V \rightarrow \R^2$, where $F(v)$ is the location of the node labeled $v$, and $\R^2$ is endowed with Euclidean distance -- we refer to this as the \emph{canonical node feature function}. On the other hand, defining a meaningful adjacency function for a shape graph requires more care. The basic idea is to model $G$ as an attributed graph over $V$ with values in the space of plane curves; i.e., if nodes $v_i$ and $v_j$ are endpoints of a curve $C_k$, then we wish to define our adjacency function as $A(v_i,v_j) = C_k$. This leads to the problem of defining an efficiently computable geodesic metric on the space of plane curves. We now describe the details of such a metric.

Formally, an \emph{unparameterized plane curve} is a smooth  immersion $\beta:[0,1] \to \R^2$ considered up to the equivalence relation $\beta \sim \beta \circ \gamma$ for any $\gamma \in \Gamma$, where $\Gamma$ is the group of diffeomorphisms of $[0,1]$. We moreover identify curves up to rigid translation---that is, $\beta \sim \beta + z$ for any constant translation vector $z \in \R^2$. It will be convenient to represent a (rigid translation class of a) parameterized curve $\beta: [0,1] \to \real^2$ by its \emph{square-root velocity function (SRVF)} $q: [0,1] \to \real^2$, given by $q(t)= \dot{\beta}(t)/\sqrt{|\dot{\beta}(t)|}$. A diffeomorphism $\gamma \in \Gamma$ acts on an SRVF via $(\gamma,q) \mapsto (q \circ \gamma) \sqrt{\dot{\gamma}}$, and the map $\beta \mapsto q$ taking a curve to its SRVF is equivariant with respect to these two actions of $\Gamma$; that is, $\beta \circ \gamma \mapsto (q \circ \gamma) \sqrt{\dot{\gamma}}$. Let $[q]$ denote the equivalence class of $q$ under the latter action of $\Gamma$ and let  ${\cal S} = \{ [q] | q \in \ltwo([0,1],\real^2)\}$ denote the \emph{shape space}. The reason for representing curves via their SRVFs is the existence of a convenient geodesic metric \[d_{\mathrm{SRV}}([q_0], [q_1]) := \inf_{\gamma \in \Gamma} \| q_0 -  ( q_1\circ \gamma)\sqrt{\dot{\gamma}}\|_2.\] The convenience of this metric lies in the extensive literature on its theoretical properties and in the existence of efficient algorithms for its application to statistical shape analysis. For example, $d_\mathrm{SRV}$ can be efficiently and accurately approximated via dynamic programming, and if we expand the regularity class of plane curves $\beta$ under consideration to include all absolutely continuous curves, then the image of the SRVF map is all of $\mathcal{S}$---for details, we refer the reader to the recent 
monograph~\cite{srivastava2016functional} and references therein.

In our previous work~\cite{guo2020-BANs}, we represented a shape graph $G=(V,E)$ as an $\mathcal{S}$-attributed graph $A:V \times V \to \mathcal{S}$, where $\mathcal{S}$ is endowed with the $d_{\mathrm{SRV}}$ metric, according to the following conventions. If $v_i, v_j \in V$ are endpoints of a curve $C_k \in E$, we defined $A(v_i,v_j) = [q]$, where $[q]$ is an equivalence class of SRVFs representing the shape of $C_k$ (i.e., $q$ is the SRVF of a parameterization $\beta$ of $C_k$). To define the value of $A$ on nodes $v_i$ and $v_j$ which are not the endpoints of a common edge, we declared $A(v_i,v_j) = {\bf 0}$, where ${\bf 0}$ is the SRVF constantly equal to the zero vector---we refer to this as the \emph{null curve}. With this model in place, the framework described in the previous subsection defines a metric, \eqref{eqn:graph_matching_metric_edges} with $d = d_{\mathrm{SRV}}$, on the space of shape graphs (at least those shape graphs whose node sets have the same cardinality---a restriction which we will remove later), or \eqref{eqn:graph_matching_metric_edges_and_nodes} on the space of shape graphs with node features, with $D$ equal to Euclidean distance.

While the convention of representing the lack of an edge between nodes with the null curve is natural in analogy with common practice in modeling weighted (classical) graphs, it is somewhat counterintuitive from a geometric perspective. For example, if a shape graph has a very short edge joining nodes $v_i$ and $v_j$, this would be measured as being close to the same shape graph with the edge between $v_i$ and $v_j$ deleted (i.e. $d_{\mathrm{SRV}}([q],{\bf 0})$ is small when $[q]$ represents a short curve), even though such an edit amounts to a large change in graph topology. To remedy this, we now introduce a modified shape graph model.

\subsection{Weighted Shape Graph Model}

To solve the problem of choosing a geometrically intuitive value for $A(v_i,v_j)$ when nodes $v_i$ and $v_j$ are not the endpoint of a common edge, we decouple adjacency information from shape information. We now represent an edge by a pair $([q],w)$, where $[q] \in {\cal S}$ is the shape and $w \in \real_+ := \{r \in \real \mid r \geq 0\}$ is \emph{weight} or \emph{thickness} of the edge. Incorporating this extra parameter into the model allows us to indicate the presence of an edge by a positive weight. On the othe hand, if a real edge does not exist between two vertices, and we set the corresponding weight to zero, then the shape variable in the ordered pair should be irrelevant. To formalize this idea, we introduce a new quotient metric space. 

Consider the closed subspace ${\cal S} \times \{0\} \subset {\cal S} \times \R_+$. We define the \emph{space of weighted shapes} to be the quotient space ${\cal W} := ({\cal S} \times \R_+)/({\cal S} \times \{0\})$---that is, ${\cal W}$ contains shapes with weights, but any two shapes with weight zero are considered to be equivalent. By abuse of notation, we denote elements of ${\cal W}$ as $([q],w)$, with the understanding that, in ${\cal W}$, $([q_0],0)$ and $([q_1],0)$ are identified for any two shapes $[q_0]$ and $[q_1]$. The equivalence class of $([q],w)$ in ${\cal W}$ is referred to as a \emph{weighted shape}.  

\begin{definition}
A \emph{weighted shape graph over node set $V$} is represented by a pair $(A,F)$, where $A$ is a $\mathcal{W}$-valued adjacency function $A: V \times V \to \mathcal{W}$ and $F$ is an $\R^2$-valued node feature function $F:V \to \R^2$.
\end{definition}

To represent a shape graph $G = (V = \{v_i\}_{i=1}^n, E = \{C_i\}_{i=1}^m)$ in this formalism, we use the canonical node feature function $F$ and we define $A(v_i,v_j)$ to be the $\mathcal{W}$-equivalence class of $([q],w)$, where 
\begin{itemize}
    \item $w=0$ if and only if $v_i$ and $v_j$ are not the endpoints of a  curve in $E$---in this case, the value of the shape variable $[q]$ is irrelevant and we can set it to ${\bf 0}$ without loss of generality;
    \item if $v_i$ and $v_j$ are the endpoints of a curve $C_k \in E$, we take  $q$ to be the SRVF for any parameterization of $C_k$. This leaves some leeway to define the weight $w$ in a data-dependent manner. A natural choice is to set $w=1$ in this case, but we discuss below alternative strategies that give better results empirically.
\end{itemize}
Under this modeling choice, nodes joined by a short edge are well distinguished from unconnected nodes, thus alleviating the issue raised in the last subsection. It remains to define a geodesic metric on $\mathcal{W}$ which gives intuitive geodesics between $\mathcal{W}$-attributed shape graphs. 

We now define a one parameter family of metrics on ${\cal W}$. Let $([q_j],w_j)$, $j \in \{0,1\}$ be a pair of weighted shapes and let $\eta > 0$ be a balance parameter. We define a function $d_\eta:{\cal W} \times {\cal W} \to \real$ by
\begin{align}
    &d_\eta(([q_0],w_0),([q_1],w_1)) \nonumber \\
    &\qquad  = \min \{d_\mathrm{SRV}([q_0],[q_1]) + \eta \lvert w_0 - w_1 \rvert, \eta(w_0 + w_1)\}. \label{eqn:weighted_shapes_metric}
\end{align}
The next result says that $d_\eta$ defines a metric with intuitive geometric properties in the context of shape graph modeling. The proof is provided in the appendix.

\begin{theorem}\label{thm:metric}
For all $\eta > 0$, $d_\eta$ is a well-defined, geodesic metric on $\mathcal{W}$ such that for all $([q_i],w_i) \in {\cal W}$, $i \in \{0,1\}$:
\begin{enumerate}
    \item If $w_0, w_1 > 0$ then, for sufficiently large $\eta$, \[d_{\eta}\big(([q_0],w_0),([q_1],w_1)\big) = d_{\mathrm{SRV}}([q_0],[q_1]) + \eta |w_0 - w_1|\] and a geodesic joining $([q_0],w_0)$ to $([q_1],w_1)$ is given by $([q_{u}],w_u)$, where $[q_{u}]$ is the SRV geodesic from $[q_0]$ to $[q_1]$ and $w_u$ is a linear interpolation from $w_0$ to $w_1$;
    \item If $w_0 > 0$ then \[d_{\eta} \big(([q_0],w_0),([q_1],0)\big) = \eta\] and a geodesic joining $([q_0],w_0)$ to $([q_1],0)$ is given by $([q_{u}],w_u)$, where $[q_{u}]$ is constantly $[q_0]$ (utilizing the equivalence $([q_1],0) \sim ([q_0],0)$) and $w_u$ is a linear interpolation from $w_0$ to $0$;
    \item $d_{\eta}\big(([q_0],0),([q_1],0)\big) = 0$.
\end{enumerate}
\end{theorem}

We now consider weighted shape graphs with adjacency functions $A: V \times V \to \mathcal{W}$ as $\mathcal{W}$-attribued graphs, where $\mathcal{W}$ is endowed with the metric $d_\eta$ for some fixed choice of $\eta > 0$. 

Geodesics between weighted shape graphs, in the sense of \eqref{eqn:graph_matching_metric_edges} with $d = d_\eta$, admit intuitive visualizations. Consider shape graphs $G_i = (V_i,E_i)$ for $i=0,1$, where we assume for simplicity that $|V_0| = |V_1| = n$ and write $V_0 = \{v^0_i\}_{i=1}^n$ and $V_1 = \{v^1_i\}_{i=1}^n$ (the case of different node set cardinalities will be handled below). Let $\sigma \in S_n$ be a permutation realizing the minimum \eqref{eqn:graph_matching_metric_edges}. We visualize the geodesic between the associated weighted shape graphs $G_0$ and $G_1$ by describing its behavior on edges. We use an example of a shape graph geodesic in Figure~\ref{fig:geodesics_examples1} to illustrate the following points: 
\begin{figure*}[t]
        \centering
        \includegraphics[height=1.2in]{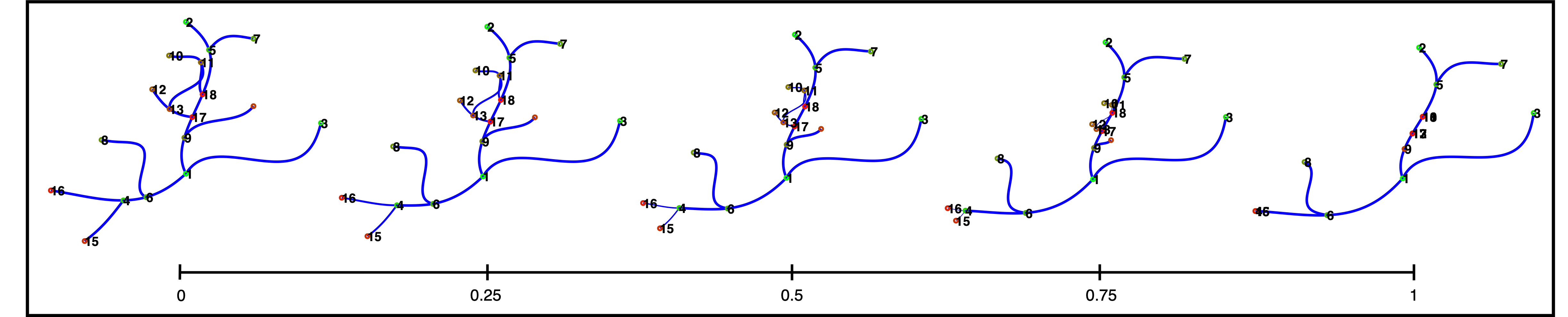}
        \caption{Geodesic between a simulated graph pair for visualizing the growing and shrinking of edges.}
        \label{fig:geodesics_examples1}
\end{figure*}
\begin{itemize}
    \item if $v^0_i$ and $v^0_j$ are endpoints of a common edge and $v^1_{\sigma(i)}$ and $v^1_{\sigma(j)}$ are as well, then the edge visualization follows the $d_{\mathrm{SRV}}$ geodesic as the shape graphs are interpolated (cf.\ Property (1) of Theorem~\ref{thm:metric});
    \item if $v^0_i$ and $v^0_j$ are endpoints of a common edge, but  $v^1_{\sigma(i)}$ and $v^1_{\sigma(j)}$ are not, then the weight of the edge interpolates to zero along the geodesic, and this is visualized by treating the weight as an opacity parameter---intuitively the edge in the first graph smoothly vanishes along the geodesic interpolation (cf. Property (2) of Theorem~\ref{thm:metric});
    \item the visualization in the previous scenario can be embellished if either $v^0_i$ or $v^0_j$ is a \emph{leaf} of the shape graph (i.e., it is the endpoint of a single edge), in which case we can instead visualize the edge geodesic as shrinking the edge to a point;
    \item if $v^0_i$ and $v^0_j$ are not endpoints of a common edge, but  $v^1_{\sigma(i)}$ and $v^1_{\sigma(j)}$ are, then the opposite of one of the two previous conventions is taken, i.e., the edge smoothly phases into existence over the course of the geodesic interpolation.
\end{itemize}
Observe that these behaviors agree with the geodesic behaviors described in Theorem~\ref{thm:metric}. We note that similar visualization conventions were used in our previous work~\cite{guo2020-BANs} (thinning and thickening of edges), but this lacked a precise justification in terms of the metric, and this is a motivation for introducing the new weighted shape graph formalism in the current paper.

To summarize this development, the metric $d_{\eta}$ derived here is used for the edge metric $d$ in Eqn.~\ref{eqn:graph_matching_metric_edges_and_nodes}. For the node distance $D$ we will use the Euclidean distance between the node feature vectors. 
Further technical details of metric and geodesic computation are given below. We close this subsection with a remark on previous related work.

\begin{remark}
The shape graph framework of~\cite{Sukurdeep-arXiv-2021} also includes weights on the edges, but the role played by the weights is to facilitate partial matching between graphs, rather than interpretable geodesics and improved node matching. Indeed, the node matching procedure of~\cite{Sukurdeep-arXiv-2021} is quite different from ours and is derived from variational principles. Application of our weighted shape graph model to partial  matching will be a future research direction.
\end{remark}

\subsection{Computing Geodesics}\label{sec:computational_details}

In this subsection, we collect several technical details involved in computing geodesics between shape graphs and provide some examples illustrating the algorithm.

\subsubsection{Different node cardinalities.}\label{sec:node_cardinalities} In practice, we typically wish to compare pairs of shape graphs $G_i = (V_i,E_i)$, $i \in \{0,1\}$, whose node sets have different cardinalities, $|V_0| \neq |V_1|$. We handle this by augmenting each graph, as follows. Let $A_i:V_i \times V_i \to \mathcal{W}$ denote the adjacency function for $G_i$. We replace $A_i$ with $\widetilde{A}_i: \widetilde{V} \times \widetilde{V} \to \mathcal{W}$, and $F_i$ with $\widetilde{F}_i: \widetilde{V} \to \mathbb{R}^2 \cup \{\star\}$ ($\star$ is an abstract point appended to $\R^2$), where $\widetilde{V} = V_0 \cup V_1$ and 
\[
\widetilde{A}_i(v_j,v_k) = \left\{\begin{array}{cc}
A_i(v_j,v_k) & \mbox{ if $v_j,v_k \in V_i$} \\
({\bf 0},0) & \mbox{ otherwise.}
\end{array}\right.
\] 
\[
\widetilde{F}_i(v_j) = \left\{\begin{array}{cc}
F_i(v_j) & \mbox{ if $v_j \in V_i$} \\
\star & \mbox{ otherwise.}
\end{array}\right.
\]
Intuitively, we are padding the shape graphs with abstract nodes that have no interaction with the original, ``true" nodes. We refer to the additional nodes as \emph{null nodes}. Since the augmented weighted shape graphs are defined over vertex sets of the same cardinality, this puts us back in the setting of the previous subsection. In the following, we will assume without loss of generality that all pairs of shape graphs have node sets of equal cardinality.

\subsubsection{Solving the shape graph registration problem.} An obvious question which needs to be addressed is how the optimization problems \eqref{eqn:graph_matching_metric_edges} and \eqref{eqn:graph_matching_metric_edges_and_nodes} are (approximately) solved in the shape graph setting. We first remark on the complexity of solving \eqref{eqn:graph_matching_metric_edges} in a simpler context. Suppose that $A_i:V_i \times V_i \to \{0,1\}$, $i \in \{0,1\}$, are the adjacency functions of graphs with binary attributes and the target metric space $\{0,1\}$ is endowed with the obvious metric---that is, the $A_i$ are adjacency matrices in the sense of classical graph theory. Even in this (as simple as possible) setting, the optimization problem \eqref{eqn:graph_matching_metric_edges} is notoriously difficult---it is called the \emph{graph matching problem}~\cite{cour2007balanced,leordeanu2005spectral,
leordeanu2009integer,zanfir2018deep,zhou2015factorized}, an instance of a quadratic assignment problem which is well known to be NP-Hard~\cite{johnson1979computers}. Our problem is more complicated than graph matching as we also seek registration of points on edges across shape graphs. We now describe a heuristic algorithm for computationally tractable approximation of solutions to \eqref{eqn:graph_matching_metric_edges} and \eqref{eqn:graph_matching_metric_edges_and_nodes} for shape graphs.

We first consider the graph matching problem \eqref{eqn:graph_matching_metric_edges}, which does not yet incorporate node feature functions. Let $A_i:V_i \times V_i \to \mathcal{W}$, $i \in \{0,1\}$, be weighted shape graph adjacency functions and assume (without loss of generality, according to the discussion in the previous subsection) that $|V_0| = |V_1| = n$. The graph matching problem \eqref{eqn:graph_matching_metric_edges} (with $d = d_\eta$) is replaced by the problem 
\begin{equation}\label{eqn:graph_matching_with_affinity}
    \max_{P \in \mathcal{P}} \mathrm{vec}(P)^T K \mathrm{vec}(P),
\end{equation}
where $\mathcal{P}$ is the set of $n \times n$ permutation matrices, $\mathrm{vec}(P) \in \R^{n^2}$ is the column vector obtained by stacking the columns of $P$ and $K \in \R^{n^2 \times n^2}$ is an \emph{affinity matrix}, defined in terms of $A_0$ and $A_1$ as follows. The affinity matrix is indexed as $K = (k_{ajbjk})_{a,j,b,k = 1}^n$, with $v^0_a,v^0_b \in V_0$ and $v^1_j,v^1_k \in V_1$. The entries are given by
\begin{equation}\label{eqn:affinities}
k_{ajbk}=\lambda\left(1-\frac{d_\eta(([q^0_{ab}],w^0_{ab}),([q^1_{jk}],w^1_{jk}))}{\max_{c,\ell,d,m} d_\eta(([q^0_{cd}],w^0_{cd}), ([q^1_{\ell m}],w^1_{\ell m}) )}\right),
\end{equation}
where $A_0(v^0_a,v^0_b) = ([q^0_{ab}],w^0_{ab})$ and $A_1(v^1_j,v^1_k) = ([q^1_{jk}],w^1_{jk})$, provided $w^0_{ab},w^1_{jk} > 0$ (real edges).
Thus, if the distance between $([q^0_{ab}],w^0_{ab})$ and $([q^1_{jk}],w^1_{jk})$ is close to zero, the affinity matix encourages $v^0_{a}$ to match to $v^1_j$ and $v^0_b$ to match to $v^1_k$, whereas matching is discouraged when the distance is close to maximum.  

The reason for replacing the graph matching \eqref{eqn:graph_matching_metric_edges} with \eqref{eqn:graph_matching_with_affinity} is that the latter can be efficiently approximated via the Factorized Graph Matching (FGM) algorithm~\cite{zhou2015factorized}, which makes clever use of the sparsity structure of $K$ in order to reduce computational complexity (see Section~\ref{subsec:computational_complexity}). We remark that the problems \eqref{eqn:graph_matching_metric_edges} and \eqref{eqn:graph_matching_with_affinity} are not exactly equivalent, but a solution to \eqref{eqn:graph_matching_with_affinity} gives a heuristic upper bound for the true problem \eqref{eqn:graph_matching_metric_edges}. More refined approaches to approximating \eqref{eqn:graph_matching_metric_edges}, especially using graph neural-networks, are the subject of ongoing research.

\subsubsection{Node attributed graph matching.} We now describe our method for approximating shape graph distance with node feature functions incorporated; i.e., our approximation of the node attributed graph matching problem~\eqref{eqn:graph_matching_metric_edges_and_nodes}. Let us assume that our shape graphs have no {\emph self-loops}, so that the endpoints of any curve in a shape graph are distinct. Then, for any pair of shape graphs, the affinity matrix $K$ defined above has zeros along its diagonal: $k_{ajaj} = 0$.

Recall that our vertex sets have been padded with abstract null nodes (Section~\ref{sec:node_cardinalities}), which do not have a fixed physical location and that we extended our node feature functions to functions $\widetilde{F}_i:\widetilde{V}_i \to \R^2 \cup \{\star\}$, where $\star$ is an abstract point appended to $\R^2$. We extend Euclidean distance on $\R^2$ to a divergence on the extended plane $\R^2 \cup \{\star\}$ by setting $D(x,y) = \|x-y\|$ if $x,y \in \R^2$, $D(\star,\star) = 0$ and $D(x,\star) = D(\star,x) = E$ for $x \in \R^2$, where $E > 0$ is a tunable parameter. Using this as the node attribute distance and $d_\eta$ as the edge attribue distance in \eqref{eqn:graph_matching_metric_edges_and_nodes} yields our final shape distance.

We now describe our method for incorporating node features into the FGM optimization scheme to approximate soultions of \eqref{eqn:graph_matching_metric_edges_and_nodes}. The FGM algorithm allows one to incorporate edge-agnostic node affinities into the graph matching problem by introducing diagonal elements. If $v^0_a$ and $v^1_j$ are null nodes, our methodology sets $k_{ajaj} = (1-\lambda)$ -- recall that $\lambda \in [0,1]$ is a parameter used to balance the roles of edge and node features in the distance computation; thus null nodes always have the same matching affinity. If $v^0_a$ and $v^1_j$ are real nodes, we set 
\[
k_{ajaj} = (1-\lambda)\left(1-\frac{D(\widetilde{F}_0(v^0_a),\widetilde{F}_1(v^1_j))}{\max_{b,k} D(\widetilde{F}_0(v^0_b),\widetilde{F}_1(v^1_k))}\right),
\]
so that there is a higher affinity if the node positions are close to one another. Finally, if $v^0_a$ is a real node and $v^1_j$ is a null node, or vice verse, we set
\[
k_{ajaj} = (1-\lambda)\left(1- \frac{e \cdot M}{\max_{b,k} D(\widetilde{F}_0(v^0_b),\widetilde{F}_1(v^1_k))}\right).
\]
This last expression needs some explanation: 
conceptually, instead of using $E = D(\widetilde{F}_0(v^0_a),\widetilde{F}_1(v^1_j))$ in the numerator of the second term, we use an adaptive dissimilarity $e \cdot M$, where $e > 0$ is a tunable parameter and $M = M(V_0,V_1)$ is an estimate of the mean Euclidean distance between the real nodes of the respective shape graphs (obtained as a sample average, for efficiency).  

Empirical evidence shows that augmenting the affinity matrix as described above provides higher quality shape comparisons.

\subsubsection{Parameter selection.} Our framework for shape graph comparison involves the following parameters:
\begin{itemize}
    \item the balance parameter $\eta$ appearing in the weighted shape graph metric $d_\eta$;
    \item the parameter $e$ controls real-to-null node affinities;
    \item the weights $\lambda$ and $(1-\lambda)$ are the relative weights on the edge and node attributes respectively.
    \item the choice of weight $w$ for each curve in a shape graph.
\end{itemize}
Once a choice of weight methodology has been made in the shape graph modeling stage, the first two parameters $\eta$ and $e$ are straightforward to tune by cross-validation. The choice of weight requires more care. One approach, as mentioned above, is to take a binary weighting scheme: real edges receive weight $1$ and null edges receive weight $0$. We have found experimentally that the graph matching algorithm performs better when we use an alternative weighting scheme: a real edge is weighted by its length and null edges receive weight $0$. For the rest of the paper, our computational results use the strategy of assigning weights by edge length and we set parameters $\eta = 1$, and $\lambda=0.5$ for all experiments and $e$ is selected from the set $\{0.5,0.7,0.9\}$ according to the experiment. 

For the rest of the paper, we generically use $d_{\mathrm{graph}}$ to refer to the distance \eqref{eqn:graph_matching_metric_edges_and_nodes} on the space of shape graphs, modeled as pairs of functions $(A,F)$ as described above, with some fixed choices of parameters $\eta, e, \lambda$ and weighting scheme. 

\subsubsection{Examples.}\label{subsec:examples}

\begin{figure*}
        \centering
        \includegraphics[scale=0.4]{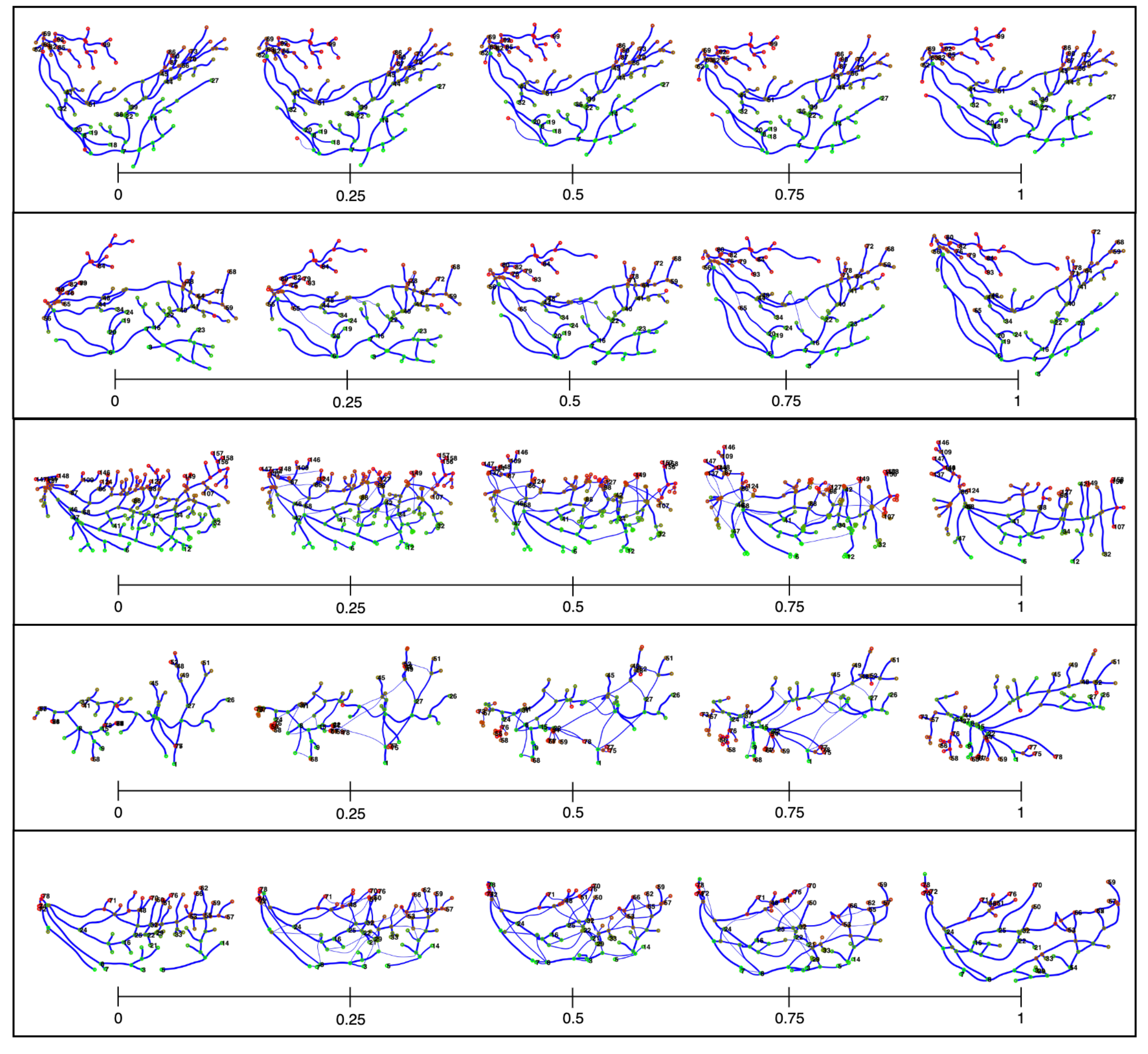}
        \caption{Geodesics between graph pairs: (a) Top 2 panels show geodesics between three simulated shape graph pairs and (b) Bottom 3 panels show real RBV graph pairs.}
        \label{fig:geodesics_examples2}
\end{figure*}

We present some examples of geodesics computed between two graphs, using some simulated shape graph pairs and some real graphs pairs from STARE~\cite{hoover2000locating,STARE-project} and DRIVE~\cite{DRIVE-challenge} datasets in Fig.~\ref{fig:geodesics_examples2}. To simulate shape graph pairs, we take a real graph and perturb its edges and nodes. We stretch the edges, remove some edges and nodes, and randomly reorder nodes. Since the original node registrations for simulated pairs are available, we treat them as ground truth and compare our results with them. Our shape-matching framework performs well on these simulated examples, with the resulting shape distances falling very close (within $0.001$) to the shape distances associated with the ground truth registrations. The geodesic deformations between real RBV networks look natural also despite a lack of ground truth to compare against. These RBV networks are complex structures and the quality of these geodesics underline the appropriateness of chosen metrics and the success of node registration.

\subsubsection{Computational complexity.}\label{subsec:computational_complexity}

Suppose we wish to compare two shape graphs, each with $n$ nodes (after null padding) and with $m_1, m_2$ edges, respectively. 
\begin{figure}
    \centering
    \includegraphics[scale=0.17]{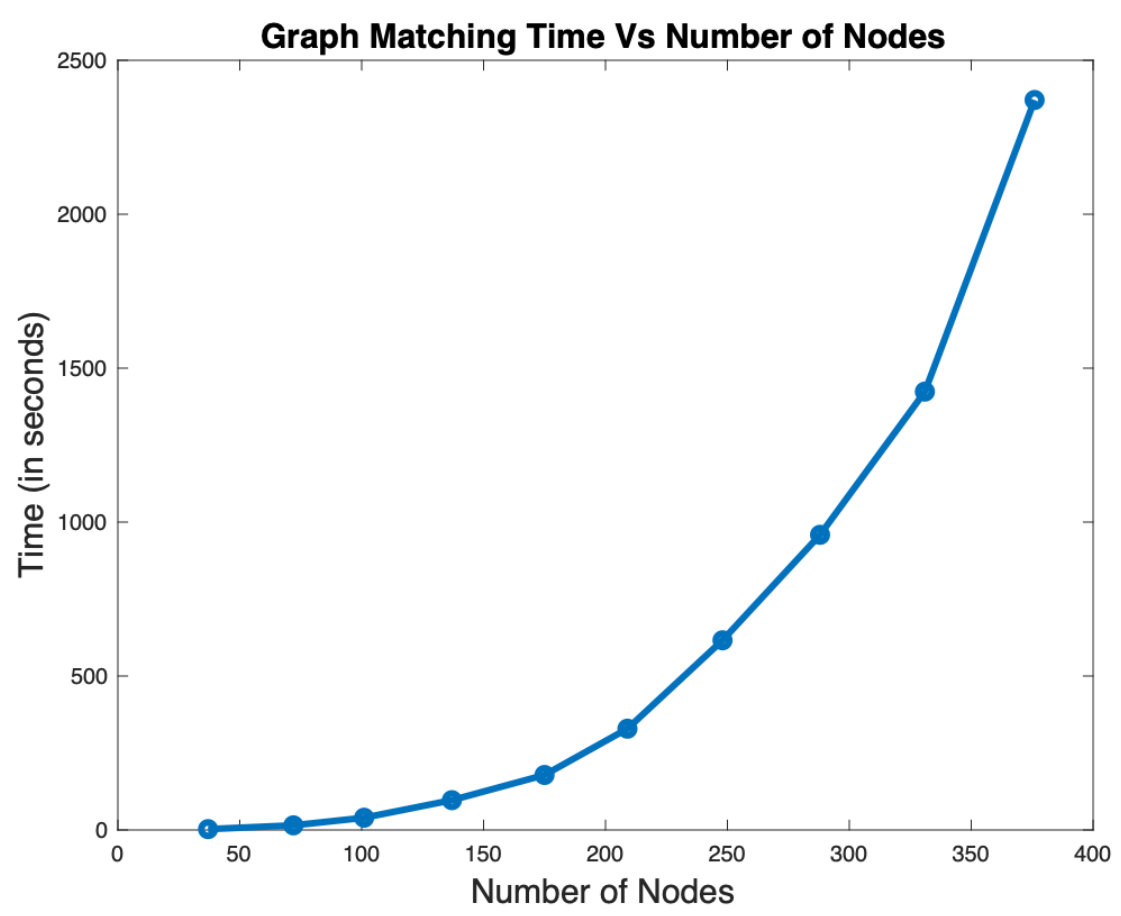}
    \caption{Time taken for graph-matching plotted versus the number of nodes in shape graphs.}
    \label{fig:time_complexity}
\end{figure}
Suppose that each edge curve is represented computationally by $\ell$ samples. We first need to construct the affinity matrix $K$. The complexity of this construction is dominated by computing SRV distances, since this involves optimization over (a discretization of) the diffeomorphism group. We use a dynamic programming algorithm~\cite{srivastava2016functional}, which has complexity $O(\ell^2)$ for each distance computation. Therefore, the overall construction of $K$ takes $O((m_1^2 + m_2^2)\ell^2)$ operations. The iterative FGM algorithm has complexity $O(t(n^3 + n(m_1 + m_2) + m_1m_2) + (n+m_1)(n+m_2)^2)$, where $t$ is the number of iterations~\cite{zhou2012factorized}, so that the overall complexity of matching two shape graphs is $O(t(n^3 + n(m_1 + m_2) + m_1m_2) + (n+m_1)(n+m_2)^2 + (m_1^2 + m_2^2)\ell^2)$. Assuming sparse graphs with $m_1,m_2$ on the order of $n$ (a realistic assumption for our data, which is tree-like), this scales as $O(tn^3 + n^2 \ell^2)$. Figure~\ref{fig:time_complexity} illustrates the time complexity for matching two shape graphs with increasing number of nodes ($n$), where $\ell=30$ and $t$ depends on $n$ and $\ell$ in the graph-matching algorithm. These computations were performed on an Ubuntu 18.04.6 LTS server with an Intel(R) Core(TM) i9-10940X CPU @ 3.30GHz with 125 GB RAM.

\section{Multiscale Representation and Matching}
In this section we develop a multiscale representation of shape graphs and use
it to develop efficient techniques for statistical analyses.

Analysis of complicated shape graphs with lots of elements (nodes and edges) and intricate geometrical features poses a challenge for several reasons. As the complexity of the shape graph data increases, the gradient descent-based graph matching estimation algorithm becomes more prone to settling into local minima of the matching cost. Moreover, providing meaningful visualizations of how important graph features evolve along geodesics in the shape space becomes difficult. We propose a novel method for representing shape graphs at multiple resolutions, which yields intuitive visualizations of a shape graph's key geometric and topological features at various scales. We can incorporate this tool into the shape-matching algorithm to improve numerical performance.

The first step in this pipeline is to cluster the nodes of a given graph hierarchically. Let $G = (V = \{v_i\}_{i=1}^n,E = \{C_j\}_{j=1}^m)$ be a shape graph. To cluster the node set $V$, we will use a metric $d^G$ on $V$ (note that this is an \emph{internal} metric for $G$, as opposed to the metrics defined earlier which are used to compare two different graphs). There are several reasonable choices, each yielding a different result. We have implemented the clustering algorithm for three choices described below.

\begin{enumerate}
\item \emph{Euclidean distance} $d^G_\mathrm{euc}:V \times V \rightarrow \R$ is defined by
$d^G_\mathrm{euc}(v_i,v_j) = \|F(v_i) - F(v_j)\|$, where $F:V \to \R$ is the canonical node feature function and 
with $\|\cdot\|$ denoting the Euclidean norm.

\item \emph{Geodesic distance} $d^G_\mathrm{geo}:V \times V \rightarrow \R$ is defined by
$$
d^G_\mathrm{geo}(v_i,v_j) = \min_{C_{k_1},\ldots,C_{k_\ell}} \sum_{a=1}^\ell \mathrm{length}(C_{k_a}),
$$
where the minimum is taken over all edge paths from $v_i$ to $v_j$. That is, such that there is a sequence of vertices $v_{k_0}, v_{k_1}, \ldots, v_{k_\ell}$, where $v_i = v_{k_0}$ is an endpoint of $C_{k_1}$, $v_j = v_{k_\ell}$ is an endpoint of $C_{k_\ell}$ and $v_{k_a} \in C_{k_{a}} \cap C_{k_{a+1}}$ for $a = 1,\ldots,\ell-1$. 

\item \emph{Effective graph resistance} $d^G_{\mathrm{eff}}:V \times V \rightarrow \R$ is defined by

\[
d^G_{\mathrm{eff}}(v_i,v_j) = ({\bf{e}}_i-{\bf{e}}_j)^{T}Q^{-1}({\bf{e}}_i-{\bf{e}}_j),
\]
where ${\bf {e}}_x$ is a vector of length $n$ that has a one at position $x$ and zeroes elsewhere. $Q^{-1}$ is the inverse of the weighted Laplacian $Q$ given by  $Q=S-W$ where $W$ is the matrix whose element $w_{ij}$ is the reciprocal of the length of the edge connecting nodes $v_i$ and $v_j$ if they are connected and 0 otherwise and $S$ is the diagonal matrix of strengths
$S_{ii} = \sum_{j=1}^n w_{ij}$

\end{enumerate}

For a metric $d^G \in \{d^G_\mathrm{euc},d^G_\mathrm{geo}, d^G_{\mathrm{eff}}\}$, we apply a hierarchical clustering algorithm (we use complete linkage clustering, but other methods are also valid) to the metric space $(V,d^G)$, providing a linkage dendrogram for $G$. 
Given a linkage dendrogram for $G$ and a scale $h$ ranging uniformly from $0$ to $1$, we produce a representation of $G$ at resolution level $h$, denoted by $G^h \equiv (V^h,A^h,F^h)$, as follows. 
Choosing the number of clusters $k$ as a fraction ($h$) of the total number of nodes at full resolution gives a partition in the nodes $V$ of $G$ into $k$ subsets (determined by the linkage structure). Each of the $k$ clusters is represented by a node $v_i^h \in V^h$, $i \in \{1,\ldots,k\}$. We define $F^h(v^h_i)$ to be the Euclidean average of the node positions of all nodes in the cluster represented by $v^h_i$.  If there is no edge of $G$ joining a node in cluster $i$ to a node in cluster $j$, then we set $A^h(v^h_i,v^h_j) = (\textbf{0},0) \in \mathcal{W}$. Otherwise the shape of $A^h(v^h_i,v^h_j)$ is obtained by taking the Karcher mean of all intercluster edges joining cluster $i$ to $j$ in the shape space $\mathcal{S}$ (see \cite{srivastava2016functional}), then rotating and rescaling the result so that it has prescribed endpoints $F^h(v^h_i)$ and $F^h(v^h_j)$, and the weight of the edge is defined according to a predefined methodology. 

\begin{figure}
        \centering
        \includegraphics[scale=0.085]{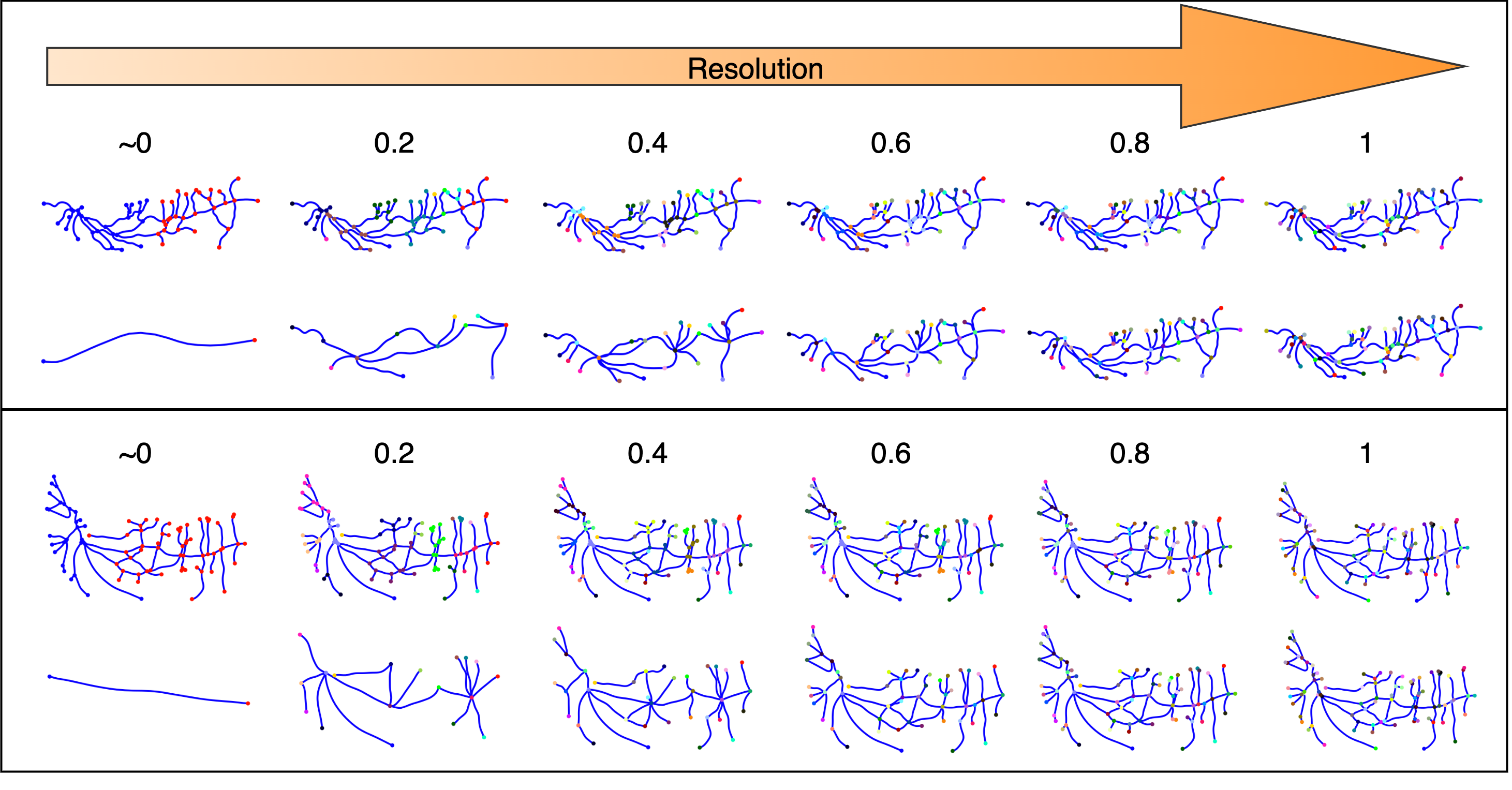}
        \caption{Multiscale representation of two RBV shape graphs using $d^G_{\mathrm{eff}}$. In each case, the top row shows node clustering (using colors) in the original graph and the bottom row shows the corresponding graphs $G^h$ at varying resolutions. }
        \label{fig:STARE44_res_clustering}
\end{figure}

We have compared the results for the three choices of $d^G$ and found that the results for $d^G_{\mathrm{eff}}$ \cite{ellens2011effective} are the best, so we present only those results here. 
Figure~\ref{fig:STARE44_res_clustering} 
shows multi-resolution representations of a RBV network with respect to $d^G_{\mathrm{eff}}$. The top row of shows the node clusters at each resolution by color and the second row shows the corresponding multi-resolution representation. We choose scale values by uniformly sampling $(0,1]$ -- this has the visual effect of allowing large scale features to be added at first, with finer scale features appearing later (corresponding to high-resolution representations).

\begin{figure}
        \centering
        \includegraphics[scale=0.095]{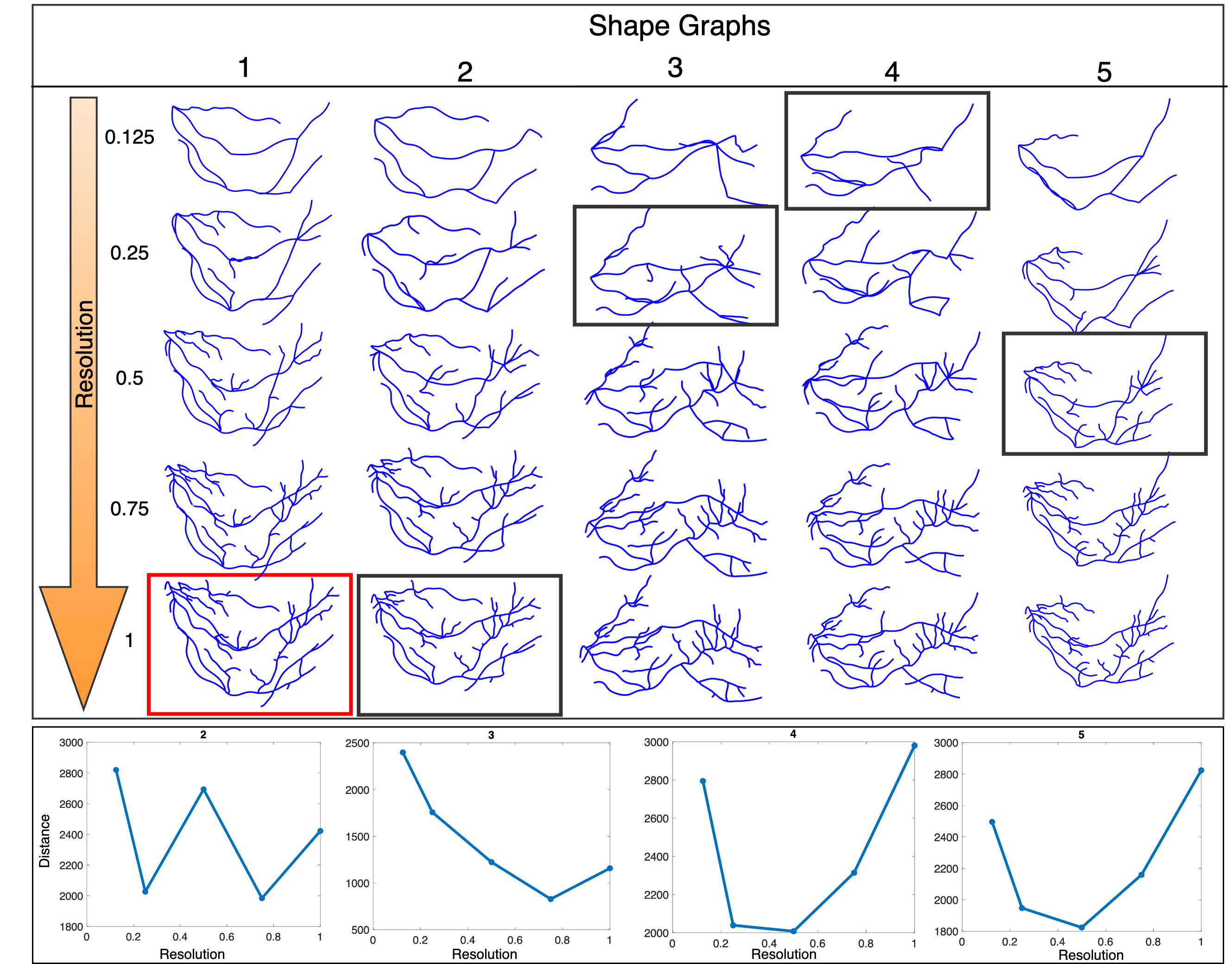}
        \caption{Each column shows multi-resolution representation of the original graph at the bottom. Taking the original graph in column 1 (with a red box) as the target, we find the closest graph in each column and display them in black box. Bottom row plots shape graph distances $d_{\mathrm{graph}}(G_1, G^{h}_i)$ versus $h$ for each column $i$, $i=2,3,4,5$.}
        \label{fig:res_across_0001}
\end{figure}

This multi-resolution representation can then be used to compare and analyze graphs with quite different complexities, as describe next. For two graphs $G_1$ and $G_2$, suppose that $G_2$ has many more node and edges than $G_1$. In order to compare them naturally, we form a multi-resolution representation $G^{h}_2$ and compute graph distance $d_{\mathrm{graph}}(G_1, G^{h}_2)$ for several levels $h$. We minimize this distance to select $G^{h}_2$ that is similar in complexity to $G_1$. Figure~\ref{fig:res_across_0001} shows this idea using several simulated graphs. Each column here shows different resolutions of the original graphs, which are displayed in the last row of graphs. The original graphs in columns 3, 4, and 5 have much higher complexity than those in columns 1 and 2. Using the original graph in column 1 (with a red box around it) as the target, we find its closest graph in each column and put a black box around it. The bottom row shows the plots of the distances $d_{\mathrm{graph}}(G_1, G^{h}_i)$ versus $h$ for each column $i$, $i=2,3,4,5$. One can see that the selected graphs resemble the target graph in complexity both visually and metrically. 

\begin{figure}
        \centering
        \includegraphics[scale=0.095]{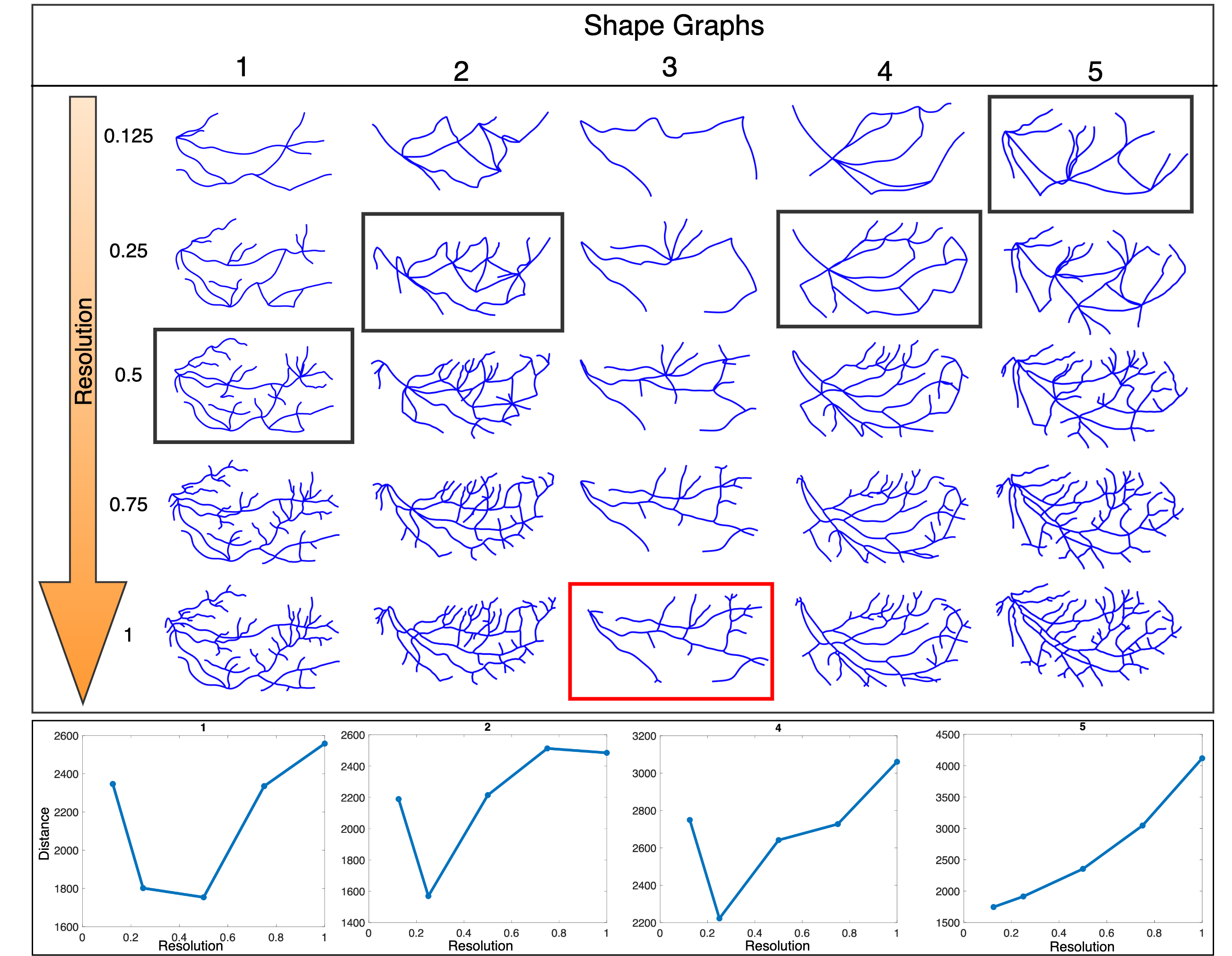}
        \caption{Same as Figure~\ref{fig:res_across_0001}.}
        \label{fig:res_across}
\end{figure}

As this example suggests, the multi-resolution representation is useful in bringing graphs to a similar resolution for more meaningful and interpretable comparisons. Graphs at similar resolution often include a comparable number of nodes and edges.  Figure~\ref{fig:res_across} shows another example of this idea using real RBV graphs. This time we use the column 3 as the target as it is relatively less complex. We observe that the closest matches to this target are also low-complexity structures. 

\begin{figure}
        \centering
        \includegraphics[scale=0.145]{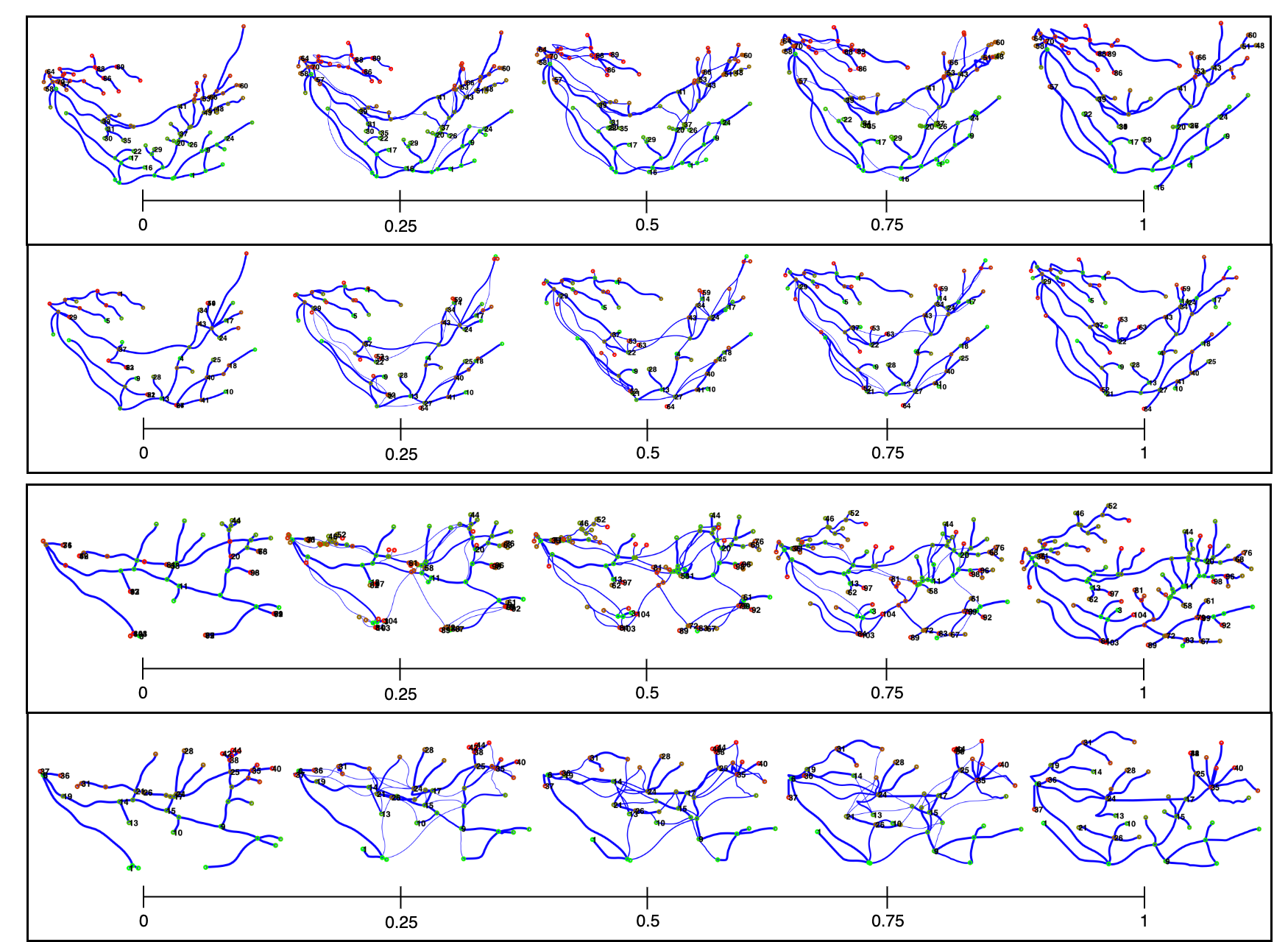}
        \caption{In each example, the top row is the geodesic between original $G_1$ and $G_2$, while the bottom row is a geodesic between $G^1$ and the simplified $G^{h^*}_2$. In the top example, $d_{graph}(G_1,G_2)=3029.1$ and $d_{graph}(G_1,G^{h^*}_2)=1823.8$. In the second example, $d_{graph}(G_1,G_2)=3329.1$ and $d_{graph}(G_1,G^{h^*}_2)=2138.3$.}
        \label{fig:geodesic_rescorr}
\end{figure}

The use of multi-resolution representation and graph distance $d_{\mathrm graph}$ helps bring graphs of different complexities to a similar level for more natural comparisons.  Given any two graphs $G_1$ and $G_2$, we can find an appropriate complexity level $h^*$ such that $G^{h^*}_2$ is matched to $G_1$. Then, we compare $G_1$ with $G^{h^*}_2$ using a geodesic path. This is illustrated using two examples in Fig.~\ref{fig:geodesic_rescorr}. In each example, the top row shows a geodesic between the original two graphs $G_1$ and $G_2$, while the second row shows the geodesic between $G_1$ and $G^{h^*}_2$, i.e., when the more complex graph has been simplified using its multi-resolution representation. One can see that by matching complexities, the geodesic deformations look more interpretable in the sense that coarser structures are deformed into coarser structures. Furthermore, the distance between the graphs at matched resolution is less than the distance between the original graphs, implying a smaller deformation in going from one to the other. 

\section{Shape Statistics}
Using multiscale representations of RBV graphs, and the metric structure imposed on the graph space, we can now perform statistical analysis 
of RBV network data. We can generate statistical summaries (compute means, medians, and 
detect outliers), perform tangent PCA in the graph space to discover dominant modes, 
perform dimension reduction to Euclidean variables, or even develop generative models for 
RBV graphs. Next, we explore some of these ideas. 
\\

\subsection{Mean and PCA of Shape Graphs}
Given a set of graph shapes $\{ G_i, i = 1,2,\dots,m\}$, we define their mean graph shape
(Fr\'{e}chet or Karcher mean) 
to be: 
$G_{\mu} = \argmin_{G \in {\cal G}} \left( \sum_{i=1}^m d_{\mathrm graph}(G,G^i)^2 \right)$, 
where $d_{\mathrm graph}$ is as defined in Eqn.~\ref{eqn:graph_matching_metric_edges_and_nodes}.
The algorithm for computing this mean shape is given in Algorithm \ref{algo:mean}. 
\begin{algorithm}
\caption{Graph Mean and TPCA  in ${\cal G}$}
\label{algo:mean}
\begin{flushleft}
Given the graphs $(A^{i}, F^i)$, $i=1,..,m$:
\end{flushleft}
\begin{algorithmic}[1]
\State Initialize an adjacency matrix $A_{\mu}$ (e.g., the largest one) and the mean node template $F_{\mu}$, to form $G_{\mu} = (A_{\mu}, F_{\mu})$.
\State Match $G^{i}$ to $G_{\mu}$ using FGM \cite{zhou2015factorized}
and SRVF~\cite{srivastava2016functional}, store the matched graph shape as $G^{i*}$, for $i = 1,.., m$.
\State Update $A_{\mu} = \frac{1}{m}\sum_{i=1}^{m} A^{i*}$ and 
$F_{\mu} = \frac{1}{m}\sum_{i=1}^{m} F^{i*}$. 
\State Repeat 2 and 3 until $\sum_{i=1}^{m} d_{\mathrm graph}(G^{i*},G_{\mu})$ convergence.

\item {\bf TPCA}: Compute shooting vectors $v_i=((A^{i*}-A_{\mu}), (F^{i*} - F_{\mu}))$ for all $i$ and perform PCA. 
Obtain directions and singular values for the principal components. 

\end{algorithmic}
\end{algorithm}

\begin{figure}
        \centering
        \includegraphics[scale=0.07]{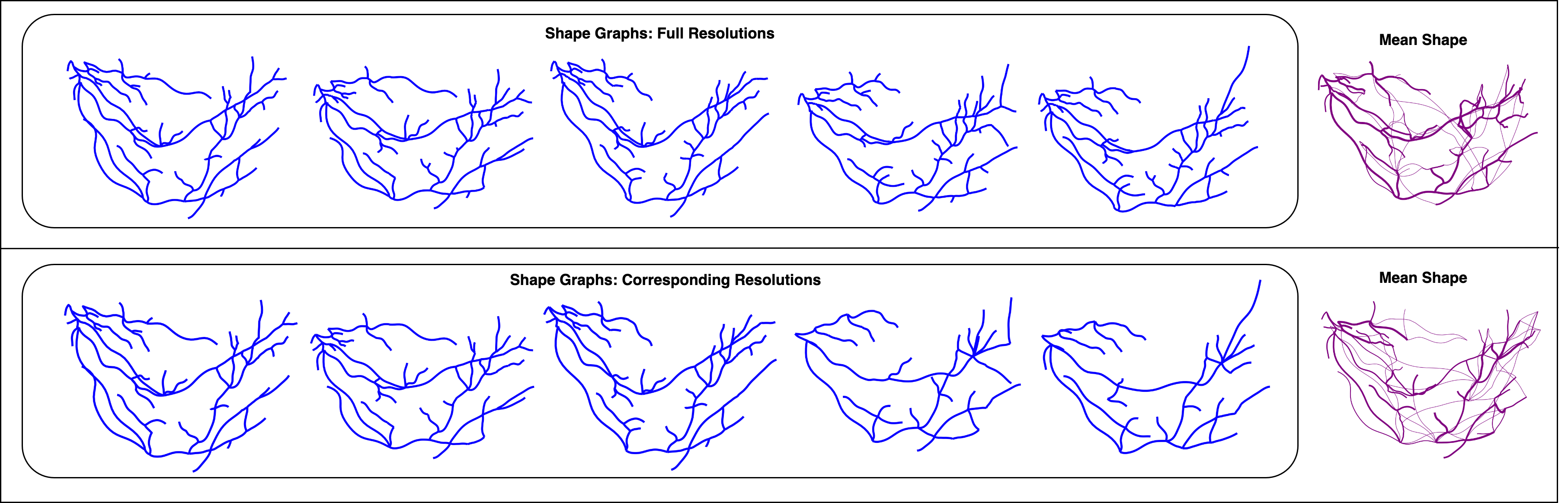}
        \caption{Plots show mean of shape graphs at full resolution (top) and corresponding resolutions  (bottom). Sum of squared $d_{graph}$ between the graphs and their mean are 1.6833e+07 and 9.3984e+06 respectively.}
        \label{fig:mean_rescorr_0001}
\end{figure}

We present an example of the mean computation in Fig. \ref{fig:mean_rescorr_0001}. In the first row we show a set of five shape graphs as given and compute their mean using Algorithm~\ref{algo:mean}. This mean shape graph is shown in the right most panel. Next, we reduce the given shape graphs to the same level of complexity and then compute the mean structure again. These results are shown in the bottom row of this figure. Note that bringing all graphs down to a lower complexity results in a cleaner mean shape graph.

Graphical shape data is often high dimensional and complex, requiring tools for 
dimension reduction for analysis and modeling. 
In past shape analysis, the tangent PCA has been used for performing 
dimension reduction and for discovering dominant modes of variability in the shape 
data. Given the graph shape metric $d_{graph}$ and the definition of shape mean $G_{\mu}$, we 
can extend TPCA to graphical shapes in a straightforward manner. 
As mentioned earlier, due to the non-registration of nodes in the raw data 
the application of TPCA directly in $\mathcal{G}$ will not be appropriate. 
Instead, one can apply TPCA in the quotient space $\mathcal{G}$, as described in the last step of Algorithm \ref{algo:mean}. 
After TPCA, graphs can be represented using low-dimensional Euclidean coefficients, which facilitates further 
statistical analysis.

\begin{figure}
        \centering
        \includegraphics[width=0.5\textwidth]{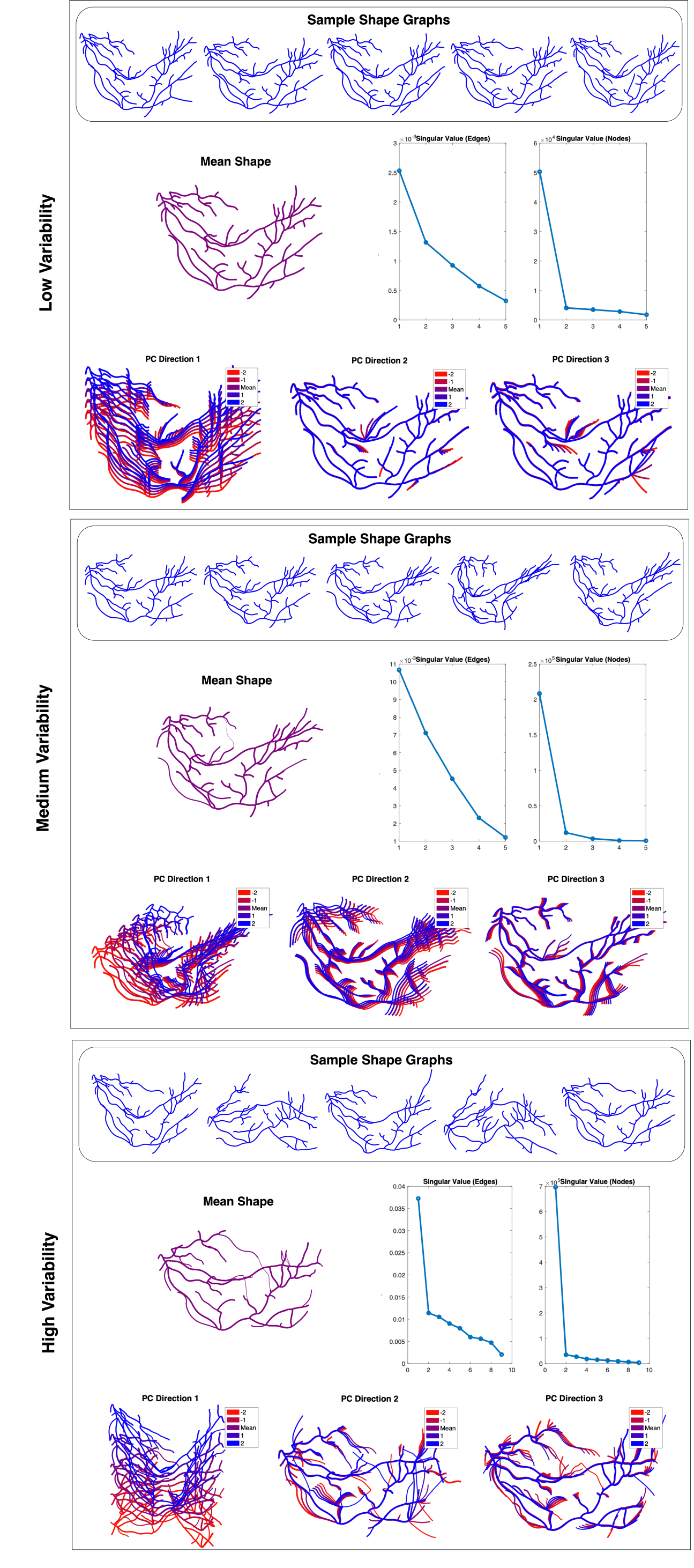}
        \caption{PCA of distorted shape graph datasets with different levels of introduced distortions.}
        \label{fig:pca}
    \end{figure}

Some examples of TPCA using simulated shape graphs is shown in Fig.~\ref{fig:pca}. This data is simulated by taking a real RBV network and distorting it several ways: bending along major axes, change edges shapes, dropping random nodes, and randomly permuting node orderings. We consider three cases where these distortions are kept small, medium, and large. The three parts of Fig.~\ref{fig:pca} show the results for these three cases. In each case, we display five random shape graphs (top left), their mean shape graph (top right), the deformations along three dominant directions (bottom left), and singular values of nodes and edges (bottom right). It is interesting to see how each of the dominant directions captures a certain kind of distortion introduced in the data during simulation. Even for the largest distortion case, where the total variation is much larger compared to the other two cases, the dominant directions provide a nice decomposition into individual effects.

\begin{figure}
        \centering
        \includegraphics[ width=0.5\textwidth]{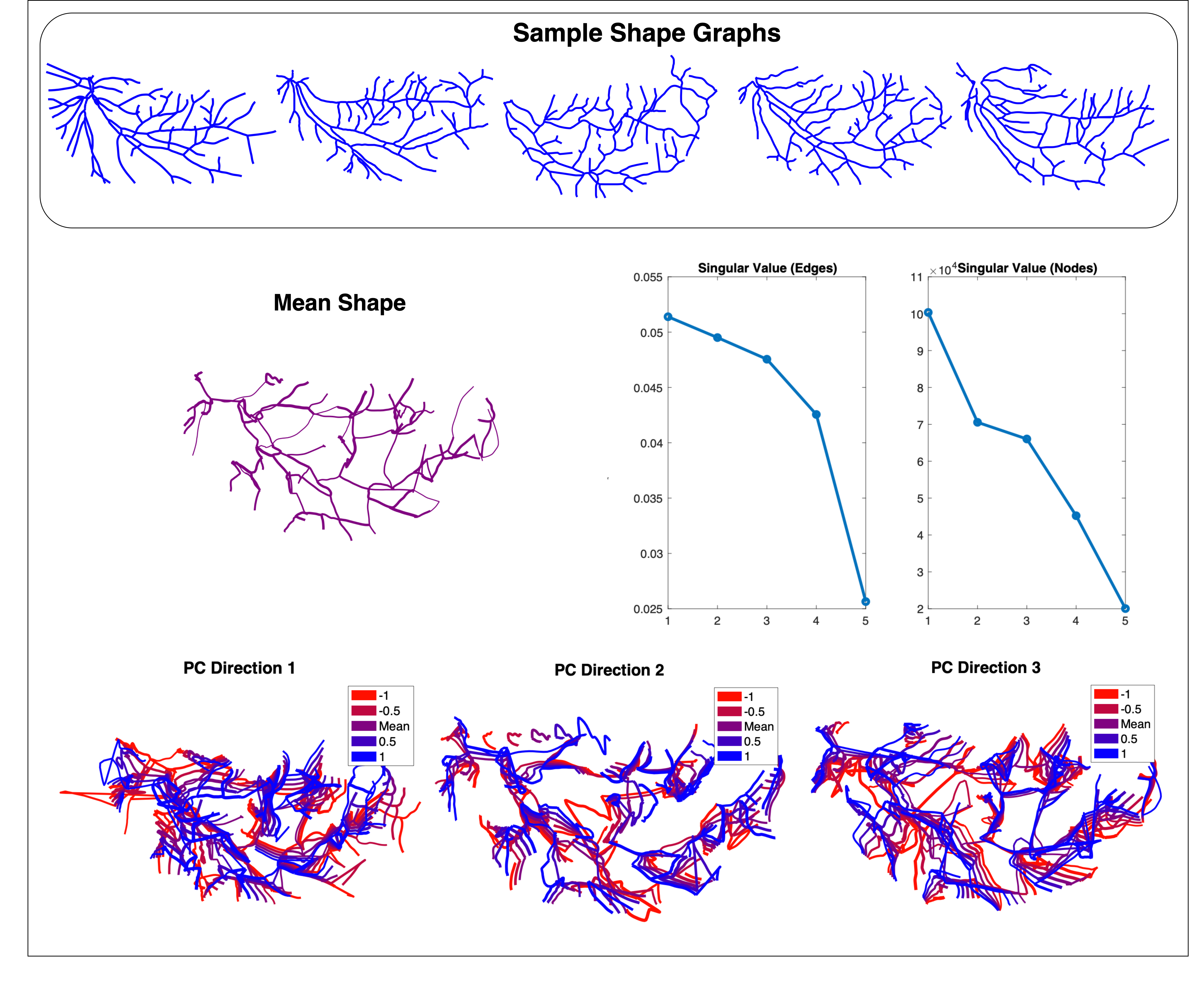}
        \caption{Example PCA ofsome shape graphs taken from the RBV dataset}
        \label{fig:pca_real}
\end{figure}

Figure~\ref{fig:pca_real} shows a similar experiment using six real RBV networks. The mean shape graph is able to capture the broad structures in the given data and the PCA also breaks the total variations into some individual distortion types. However, the decomposition is not as nice as in  the simulated because the real data feature more complex and highly correlated variability. The variability in shapes of edges is high and that shows up in the PCA plots. For a cleaner visualization of the variations along each principal direction, we truncated some of the lower weighted edges. As a result, some of the edges in PC Direction 2 appear to be disconnected but they are actually not.
    
Additional results on real RBV graphs have been presented in the supplementary material provided with the paper.

\subsection{Clustering in Shape Graphs}

In this section we illustrate the use of shape graph metric $d_{\mathrm graph}$ for clustering given shape graphs into disjoint clusters. For clustering, we will utilize a recent method proposed by Deng et al.~\cite{deng-ISBI:2022}. This takes in the matrix of pairwise distances as input and generates: (1) a set of estimated {\it modes} (significant peaks of the underlying probability density function) in the data, (2) divides the given shapes into clusters with the estimated modes at the center, and (3) finds the number of modes (or clusters) automatically from the data. It also outputs a list of detected outliers in the data. 

\begin{figure}
    \centering
    \includegraphics[scale=0.12]{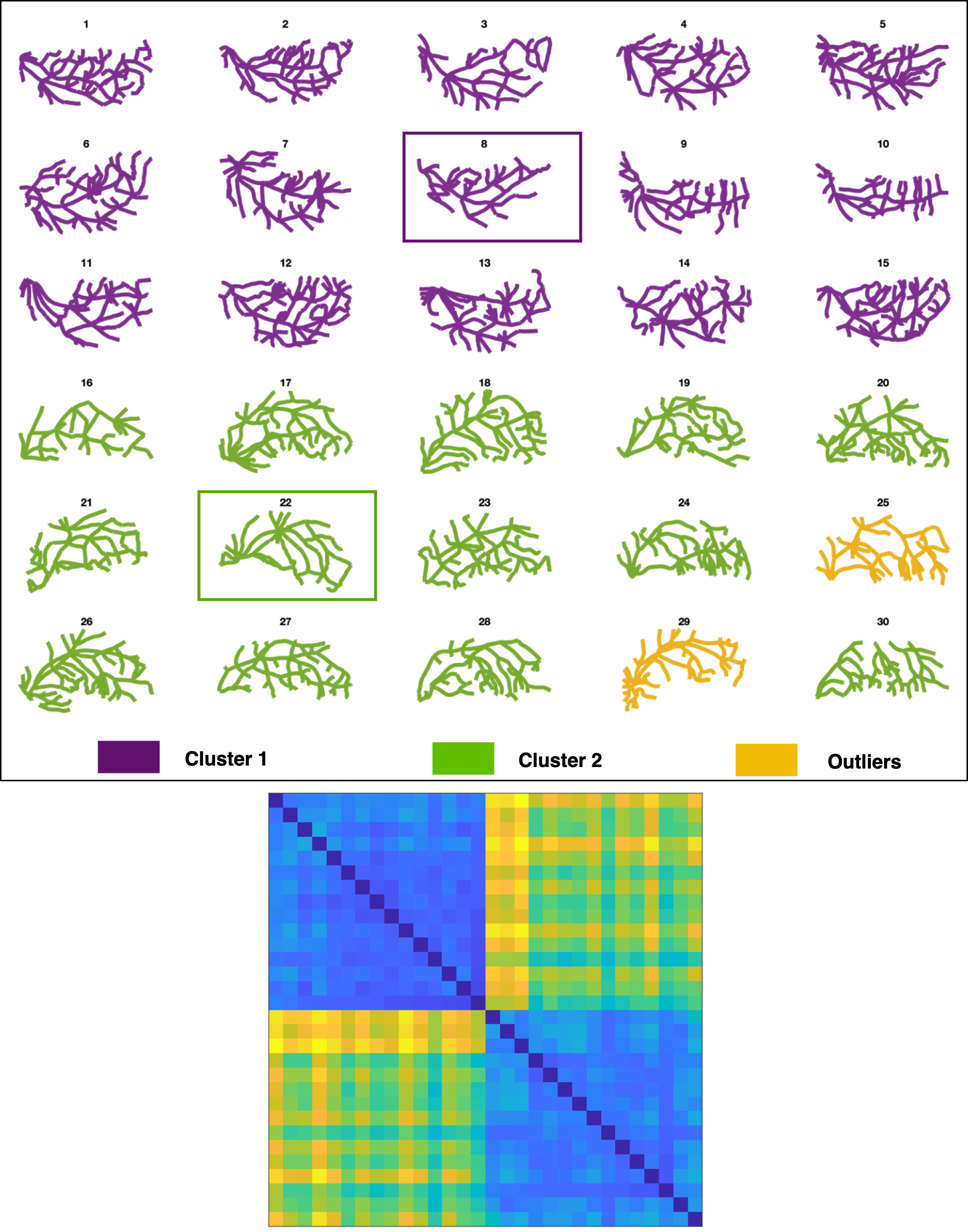}
    \caption{Clustering of shape graphs showing appropriate cluster assignment for top and bottom RBV networks. Boxed graphs are the modes for the corresponding clusters.}
    \label{fig:clustering}
\end{figure}

Consider the set of 30 shape graphs shown in the top part of Fig.~\ref{fig:clustering}. Note that the first 15 are the {\it top} parts of the full RBV networks while the next 15 are the {\it bottom} parts. We compute a $30 \times 30$ matrix of pairwise distances $d_{\mathrm graph}(G_i, G_j)$ between these graphs and use the mode-clustering algorithm of Deng et al.~\cite{deng-ISBI:2022}. The result is displayed in the bottom part of Fig.~\ref{fig:clustering}. The pairwise distance matrix, rearranged by cluster memberships, is shown in the bottom. The algorithm selected two clusters as expected, drawn in purple and green, and also labeled two shapes graphs as outliers (in yellow). The selected cluster modes are put in boxes of corresponding colors.

\section{Conclusion}
This paper develops a comprehensive framework for statistical analysis of {\it shape graphs}, graphs where edges are articulated curves characterized by their shapes. It introduces a multi-resolution representation of shape graphs that allows one to change the complexity while preserving the coarser features. 
This representation helps greatly improve the matching and comparions of shape graphs with different complexities, by bringing them to the same level. 
The paper explores the uses of these techniques for shape analysis of RBV networks. These retinal blood vessel networks have complex
structures with large variations in shapes, and have not been studied previously using formal techniques. 
This paper generates mean shapes and principal geodesics 
to discern salient patterns in RBV structures.
An important strength of 
this approach is that shape analysis is performed in the original representation space, rather than any abstract 
morphological or deep-learning feature space, and can lead to analytical generative models for such random shapes.  

There are several directions of research that follow this article. One is to develop fast and effective techniques for registration of points across shape graphs, i.e., solving Eqn.\ref{eqn:graph_matching_metric_edges_and_nodes}. Currently we use FGM but the computational cost becomes prohibitive as the number of nodes and edges become large. We are exploring an adaptation of a neural graph matching network~\cite{ngm-net:2022} to our shape registration problem. 
In another direction, one can use the mathematical representations and statistics tools in regression models for diagnosing diseases such as {\it diabetic retinopathy} or 
{\it Choroidal Neovascularization} from the observed structures.

\appendices

\section{Proof of Theorem \ref{thm:metric}}

Recall that $\W$ denotes the space of \emph{weighted shapes}, $\W = (\shapes \times \real_+)/(\shapes \times \{0\})$ and that, for $\eta > 0$, $d_\eta:\W \times \W \to \real$ is defined by
\begin{align*}
&d_\eta\big(([q_0],w_0),([q_1],w_1)\big) \\
&\qquad = \min \{d_\mathrm{SRV}([q_0],[q_1]) + \eta |w_0 - w_1|, \eta (w_0 + w_1)\},
\end{align*}
where we will continually slightly abuse notation and use $([q_j],w_j)$ to represent an equivalence class in the quotient space $\W$. 

We will first show that $d_\eta$ defines a metric. Observe that 
\[
\hat{d}_\eta\big(([q_0],w_0),([q_1],w_1)\big) := d_\mathrm{SRV}([q_0],[q_1]) + \eta |w_0 - w_1|
\]
defines a metric on $\shapes \times \real_+$---indeed, this is just a (weighted) product metric coming from the SRV metric on $\shapes$ and the standard metric on $\real_+$. The subspace $\shapes \times \{0\}$ is closed in $\shapes \times \real_+$ with respect to the metric topology. We apply the following result which follows from the discussion surrounding \cite[Proposition 1.26]{weaver2018lipschitz}.

\begin{proposition}\label{prop:quotient_metric}
Let $(X,\hat{d})$ be a metric space and let $Y \subset X$ be a closed subset. Let $[x]$ denote the equivalence class of $x \in X$ in the quotient space $X/Y$. Then 
\[
d\big([x],[x']\big) = \min\{\hat{d}(x,x'), \hat{d}(x,Y) + \hat{d}(x',Y)\},
\]
where $\hat{d}(x,Y) = \inf\{y \in Y \mid \hat{d}(x,y)\}$, is a well-defined metric on $X/Y$. 
\end{proposition}

In our setting, the distance to the subspace of interest has a closed form:
\[
\hat{d}_\eta\big(([q],w),\shapes \times \{0\}\big) = \hat{d}\big(([q],w),([q],0)\big) = \eta w.
\]
It follows that the metric construction of Proposition \ref{prop:quotient_metric} gives
\begin{align*}
&\big(([q_0],w_0),([q_1],w_1)\big) \\
& \mapsto \min\{\hat{d}\big(([q_0],w_0),([q_1],w_1)\big) , \\
&\qquad \qquad \quad \hat{d}\big(([q_0],w_0),\shapes \times \{0\} \big) + \hat{d}\big(([q_1],w_1),\shapes \times \{0\}\big) \} \\
&= \min\{\hat{d}\big(([q_0],w_0),([q_1],w_1)\big),\eta(w_0 + w_1)\} \\
&= d_\eta\big(([q_0],w_0),([q_1],w_1)\big),
\end{align*}
so that $d_\eta$ gives a well-defined metric on $\W$. 

Recall the definition of a geodesic metric space $(X,d)$ from Section~\ref{sec:metric_space_attributed_graphs}. Also recall (see, e.g. \cite[Lemma 1.3]{chowdhury2018explicit}) that to verify that a path $x_t$ in $X$ is a geodesic, it suffices to show $d(x_s,x_t) \leq (t-s)d(x_0,x_1)$. Next we show that $d_\eta$ is a geodesic metric by constructing explicit geodesics. 

Given $[q_0],[q_1] \in \shapes$, let $[q_u]$, $u \in [0,1]$ denote the SRV geodesic joining them. For $w_0,w_1 \in \real_+$, let $w_u = (1-u)w_0 + u w_1$ denote the (Euclidean) geodesic between them. Since $\hat{d}_\eta$ is simply a (weighted) product metric, for any $([q_j],w_j) \in \shapes \times \real_+$, $j \in \{0,1\}$, a geodesic joining them is given by $([q_u],w_u)$. To describe geodesics in the quotient space $\W$ with respect to $d_\eta$, it only remains to check a few cases:
\begin{enumerate}[(a)]
    \item $([q_j],w_j) \not \in \shapes \times \{0\}$ for both $j=0$ and $j=1$ and $d_\eta\big(([q_0],w_0),([q_1],w_1)\big)$ is realized by $\hat{d}_\eta\big(([q_0],w_0),([q_1],w_1)\big)$: in this case, a geodesic is given by $([q_u],w_u)$. To verify this, we will show that for all $0 \leq s \leq t \leq 1$, $d_\eta\big(([q_s],w_s),([q_t],w_t)\big)$ is realized by $\hat{d}_\eta\big(([q_s],w_s),([q_t],w_t)\big)$---the claim then follows by the fact that $([q_u],w_u)$ is a geodesic for $\hat{d}_\eta$. We have
    \begin{align*}
        &d_\eta\big(([q_s],w_s),([q_t],w_t)\big) \\
        & \quad =\min\{d_\mathrm{SRV}([q_s],[q_t]) + \eta |w_s + w_t|, \eta(w_s + w_t)\}.
    \end{align*}
    The first term in the minimum is less than or equal to $(t-s)\big(d_\mathrm{SRV}([q_0],[q_1]) + \eta |w_0 + w_1|\big)$ (because $([q_u],w_u)$ is a geodesic for $\hat{d}_\eta$), which is in turn less than or equal to $(t-s)\eta(w_0 + w_1)$ (because $d_\eta$ is realized by $\hat{d}_\eta$, by assumption). On the other hand, the second term in the minimum is equal to $\eta\big((2 - s - t)w_0 + (s+t) w_1\big)$ (by the structure of $w_u$). Finally, the claim follows from
    \begin{align*}
        0 &\leq (2-2t)w_0 + 2sw_1 \\
        &= (2-s-t)w_0 + (s+t)w_1 + (s-t)(w_0 + w_1),
    \end{align*}
    as this implies
    \begin{align*}
                & d_\mathrm{SRV}([q_s],[q_t]) + \eta |w_s + w_t|  \leq (t-s)(w_0 + w_1) \\
                & \qquad \qquad \qquad   \leq (2-s-t)w_0 + (s+t)w_1 \\
                & \qquad \qquad \qquad = \eta(w_s + w_t).
    \end{align*}
    This completes the proof of the claim for this case.

    \item $([q_j],w_j) \not \in \shapes \times \{0\}$ for both $j=0$ and $j=1$ and $d_\eta\big(([q_0],w_0),([q_1],w_1)\big)$ is realized by $\eta(w_0 + w_1)$:
    we define a geodesic piecewise. Assume, without loss of generality, that $w_0 \leq w_1$ and let $\alpha = \frac{w_0}{w_0+w_1}$; for $u \in [0,\alpha]$, consider the path $([q_0],(1-u)w_0 - u w_1)$ and for $u \in [\alpha,1]$, consider the path $([q_1],(u-1)w_0 + uw_1)$. Observe that when $u = \alpha$, the pieces of the path arrive at $([q_0],0)$ and $([q_1],0)$, respectively. Since these points are identified in $\W$, the path we have constructed is well-defined and continuous. Moreover, it is a geodesic. Indeed, if $s,t \leq \alpha$, we have that the distance between the corresponding points on along the path is given by
    \begin{align*}
        &d_\eta\big(([q_0],(1-s)w_0 - s w_1),([q_0],(1-t)w_0 - t w_1) \\
        &= \min \{\eta \lvert (1-s)w_0 - s w_1 - (1-t)w_0 + t w_1 \rvert, \\
        &\qquad \qquad \eta\big((1-t)w_0 - t w_1 + (1-t)w_0 - t w_1) \big) \\
        &\leq \eta (t-s) (w_0 + w_1) \\
        &= (t-s) d_\eta \big(([q_0],w_0),([q_1],w_1)\big),
    \end{align*}
    as required. A similar computation shows that the desired inequality holds when $s,t \geq \alpha$. Finally, suppose that $s < \alpha$ and $t \geq \alpha$. Then 
    \begin{align*}
        &d_\eta\big(([q_0],(1-s)w_0 - s w_1),([q_1],(t-1)w_0 + t w_1) \\
        &= \min \{d_\mathrm{SRV}([q_0],[q_1]) + \\
        &\qquad \qquad  \eta \lvert (1-s)w_0 - s w_1 - (t-1)w_0 - t w_1 \rvert, \\
        &\qquad \qquad \quad \eta\big( (1-s)w_0 - s w_1 + (t-1)w_0 + t w_1\big) \} \\
        &\leq \eta (t-s) (w_0 + w_1) \\
        &= (t-s) d_\eta \big(([q_0],w_0),([q_1],w_1)\big).
    \end{align*}
    
    \item $([q_j],w_j) \in \shapes \times \{0\}$ for (say) $j=0$, but not for $j=1$: In this case, a geodesic is given by $([q_1],w_u)$, with $w_u$ denoting the interpolation from $0$ to $w_1$. Observe that $([q_0],0) \sim ([q_1],0)$ in $\W$, so that this is a well-defined path. The calculation used to verify that this is a geodesic is similar to those above.
    
    \item $([q_j],w_j) \in \shapes \times \{0\}$ for both $j = 0$ and $j=1$: in this case, the points are equivalent in the quotient space, so the geodesic joining them is constant.
\end{enumerate}

Finally, we use the geodesic formulas described in (a)-(b) to prove the geodesic properties of Theorem \ref{thm:metric}. The arguments are enumerated in correspondence with the enumeration in the theorem statement:
\begin{enumerate}
    \item When $\eta \geq \frac{1}{2} d_\mathrm{SRV}([q_0],[q_1])$, we are in the situation of case (a) above, so the claim follows. In particular, taking $\eta \geq \max\{d_\mathrm{SRV}([q_0],[0]), d_\mathrm{SRV}([q_1],[0])\}$ is sufficient, by the triangle inequality.
    \item This claim follows by case (d) above.
    \item This claim follows by case (c) above.
\end{enumerate}

This completes the proof of Theorem \ref{thm:metric}.

\hfill \qedsymbol

\ifCLASSOPTIONcompsoc
  \section*{Acknowledgments}
\else
  \section*{Acknowledgment}
\fi

The authors thank Prof. Eric Klassen of FSU for helpful discussions. They also thank the creators of DRIVE and STARE datasets for making them available to the public for research. Finally, they acknowledge the NSF grants CDS\&E 1953087, NSF IIS 1955154 and NSF DMS 2107808 for supporting this research.

\ifCLASSOPTIONcaptionsoff
  \newpage
\fi

\nopagebreak 
 \bibliography{egbib}
 \bibliographystyle{plain} 
\end{document}